\documentclass[twocolumn,times,astrosymb,tighten,trackchanges]{aastex701}
\usepackage{xspace}

\newcommand{\epseri}{$\epsilon$~Eri\xspace}
\newcommand{\epserib}{$\epsilon$~Eri~b\xspace}

\newcommand{\Prot}{$P_{\rm rot}$}
\newcommand{\Teff}{$T_{\rm eff}$}

\begin{document}

\title{Worlds Next Door. III. Indirect Evidence for Enhanced Atmospheric Metallicity and/or the Presence of Water Clouds in the Nearest Jupiter-analog \epserib}
\correspondingauthor{Aniket Sanghi}

\author[0000-0002-1838-4757]{Aniket Sanghi} 
\altaffiliation{NSF Graduate Research Fellow}
\affiliation{Cahill Center for Astronomy and Astrophysics, California Institute of Technology, 1200 E. California Boulevard, MC 249-17, Pasadena, CA 91125, USA}
\email[show]{asanghi@caltech.edu}

\author[0000-0001-5864-9599]{James Mang}
\altaffiliation{NSF Graduate Research Fellow}
\affiliation{Department of Astronomy, University of Texas at Austin, Austin, TX 78712, USA}
\email{j_mang@utexas.edu}

\author[0000-0002-3414-784X]{Jorge Llop-Sayson}
\affiliation{Jet Propulsion Laboratory, California Institute of Technology, Pasadena, CA 91109, USA}
\email{jorge.llop.sayson@jpl.nasa.gov}

\author[0000-0003-2008-1488]{Eric E. Mamajek}
\affiliation{Jet Propulsion Laboratory, California Institute of Technology, Pasadena, CA 91109, USA}
\email{mamajek@jpl.nasa.gov}

\author[0000-0001-5684-4593]{William Thompson}
\affiliation{NRC Herzberg Astronomy and Astrophysics, 5071 West Saanich Road, Victoria, BC, V9E 2E7, Canada}
\email{william.thompson@nrc-cnrc.gc.ca}

\author[0000-0001-6635-5080]{Ankan Sur}
\affiliation{Department of Earth, Planetary, and Space Sciences, University of California Los Angeles, 595 Charles E Young Dr E, LA, CA 90095}
\email{ankansur@epss.ucla.edu}

\author[0000-0002-5627-5471]{Charles Beichman}
\affiliation{NASA Exoplanet Science Institute, Caltech-IPAC, Pasadena, CA 91125, USA}
\affiliation{Jet Propulsion Laboratory, California Institute of Technology, Pasadena, CA 91109, USA}
\email{chas@ipac.caltech.edu}

\author[0000-0001-5966-837X]{Geoffrey Bryden}
\affiliation{Jet Propulsion Laboratory, California Institute of Technology, Pasadena, CA 91109, USA}
\email{geoffrey.bryden@jpl.nasa.gov}

\author[0000-0001-8612-3236]{Andr\'as G\'asp\'ar}
\affiliation{Steward Observatory, University of Arizona, Tucson, AZ 85721, USA}
\email{agaspar@arizona.edu}

\author[0000-0002-0834-6140]{Jarron Leisenring}
\affiliation{Steward Observatory, University of Arizona, Tucson, AZ 85721, USA}
\email{jarronl@arizona.edu}

\author[0000-0002-8895-4735]{Dimitri Mawet}
\affiliation{Cahill Center for Astronomy and Astrophysics, California Institute of Technology, 1200 E. California Boulevard, MC 249-17, Pasadena, CA 91125, USA}
\affiliation{Jet Propulsion Laboratory, California Institute of Technology, Pasadena, CA 91109, USA}
\email{dmawet@astro.caltech.edu}

\author[0000-0002-4404-0456]{Caroline V. Morley}
\affiliation{Department of Astronomy, University of Texas at Austin, Austin, TX 78712, USA}
\email{cmorley@utexas.edu}

\author[0000-0003-2233-4821]{Jean-Baptiste Ruffio}
\affiliation{Department of Astronomy \& Astrophysics, University of California, San Diego, La Jolla, CA 92093, USA}
\email{jruffio@ucsd.edu}

\author[0000-0002-9977-8255]{Schuyler G. Wolff}
\affiliation{Steward Observatory, University of Arizona, Tucson, AZ 85721, USA}
\email{sgwolff@arizona.edu}

\author[0000-0001-7591-2731]{Marie Ygouf}
\affiliation{Jet Propulsion Laboratory, California Institute of Technology, Pasadena, CA 91109, USA}
\email{marie.ygouf@jpl.nasa.gov}

\shorttitle{JWST/NIRCam Observations of \epseri}
\shortauthors{Sanghi et al.}

\begin{abstract}

We present the most sensitive direct imaging search for the nearest ($d = 3.2$\,pc) Jupiter-analog exoplanet, \epserib, with JWST/NIRCam coronagraphy between 4--5~$\mu$m (F444W). We achieve a 5$\sigma$ contrast sensitivity $\approx3.0\times10^{-7}$ ($\Delta \approx 16.3$~mag) in the F444W filter at the expected planet separation of $\approx$1\arcsec. This is the deepest 4--5~$\mu$m contrast performance achieved for any JWST/NIRCam observation to date at these separations (and $>10\times$ better than ground-based limits). Yet, the planet remains elusive to imaging. We update the star's age to $1.1\pm0.1$~Gyr, older than previous age estimates, using the latest gyrochronology relations. This significantly impacts \epserib's inferred effective temperature ($T_{\rm eff}$), which is now expected to lie between 150--200~K based on evolutionary models for a 1~$M_{\rm Jup}$ planet. Using cloud-free Sonora Flame Skimmer models and custom PICASO patchy cloud models in the above $T_{\rm eff}$ range, we find that the F444W non-detection of \epserib can be explained by a metal-enriched atmosphere and/or an atmosphere containing water ice clouds. Both possibilities suggest that \epserib's atmosphere is strikingly similar to that of Jupiter in our Solar System. Alternatively, if we do not enforce the dynamical mass ($0.98 \pm 0.09\;M_{\rm Jup}$), a solar metallicity, cloud-free, $\lesssim0.81\;M_{\rm Jup}$ planet would be consistent with the NIRCam upper limit based on the Sonora Flame Skimmer evolutionary models. Finally, we place limits on the size of a potential ring system using the NIRCam/F210M data and discuss the opportunity to directly image \epserib with additional JWST observations, the Roman Coronagraph Instrument, the ExtraSolar Coronagraph on the Lazuli Observatory, and EELT/METIS.
\end{abstract}

\keywords{\uat{James Webb Space Telescope}{2291} --- \uat{Coronagraphic imaging}{313} --- \uat{Extrasolar gaseous giant planets}{509} --- \uat{Exoplanet Atmospheres}{487} --- \uat{Exoplanet evolution}{491} --- \uat{Exoplanet rings}{494}}

\section{Introduction} 
$\epsilon$~Eridani (\epseri) is a nearby (3.2 pc), adolescent ($1.1\pm0.1$ Gyr; this work, \S\ref{sec:sys-prop}), debris disk- and giant planet-hosting 0.8~$M_\odot$ K2V star \citep{Gray2006, gonzalez_parent_2010}. It is reminiscent of a younger version of the Solar System and, as the ninth closest star system to the Sun, serves as a benchmark for detailed characterization studies.

\epseri is a member of the original IRAS ``Fab Four" stars, with a prominent debris disk first identified as a far-infrared ($\lambda > 25\;\mu$m) excess that has since been targeted extensively \citep{aumann_iras_1985, 1986ASSL..124...61G, 1993prpl.conf.1253B, backman_epsilon_2009, greaves_extreme_2014, macgregor_epsilon_2015, chavez-dagostino_early_2016, su_inner_2017}. Atacama Large Millimeter/ submillimeter Array (ALMA) imaging found a narrow outer ring, analogous to the Kuiper Belt, centered at 69~au with a width of $12 \pm 1$~au, an inclination with respect to the sky plane of $33.7^\circ \pm 0.5^\circ$, and a position angle of $-1.1^\circ \pm 1.0^\circ$ measured anticlockwise from North  \citep{booth_northern_2017, booth_clumpy_2023}. Nulling interferometry at 10~$\mu$m obtained a strong exozodiacal dust detection \citep[$\sim$300~zodis, HOSTS Survey;][]{ertel_hosts_2020} which is seen as radial point spread function (PSF) residuals in mid-infrared direct imaging with JWST \citep{wolff_jwstmiri_2025}. The JWST 15--25.5~$\mu$m imaging further revealed a smooth dust distribution, with no evidence of sculpting by massive planets outside of 5~au, and found that the disk morphology is dominated by the inward migration of dust, produced in the outer belt, via stellar wind drag \citep{wolff_jwstmiri_2025}.

\begin{deluxetable}{lcc}
    \tabletypesize{\footnotesize}
    \tablecaption{\label{tab:prop}Stellar Properties of \epseri}
    \tablehead{\colhead{Property} & \colhead{Value} & \colhead{Ref.}}
    \startdata
    $\alpha$\tablenotemark{\scriptsize a} ($^\circ$) & $53.22829341518$ & 1\\
    $\delta$\tablenotemark{\scriptsize a} ($^\circ$) & $-9.45816821629$ & 1\\
    $\mu_{\alpha^*}$\tablenotemark{\scriptsize b} (mas\,$\mathrm{yr^{-1}}$) & $-974.758\pm0.160$ & 1\\
    $\mu_{\delta}$ (mas\,$\mathrm{yr^{-1}}$) & $20.876\pm0.120$   & 1\\
    $\varpi$ (mas) & $310.5773\pm0.1355$ & 1\\
    Distance (pc) & $3.21978^{+0.00134}_{-0.00147}$ & 2\\
    SpT & K2V(k) & 3\\
    $G$ & $3.466\pm0.003$ & 1\\
    $B_p - R_p$ (mag) & 1.140298 & 1\\
    $B_p - G$ (mag)   & 0.516696 & 1\\
    $G - R_p$ (mag)   & 0.623602 & 1\\	  	
    $B-V$ (mag)       & $0.882\pm0.007$ & 4\\
    $V$ (mag)         & $3.726\pm0.010$ & 4\\
    $J$ (mag)         & $2.175\pm0.007$ & X, 10\\
    $H$ (mag)         & $1.747\pm0.011$ & X, 10\\
    $K_s$ (mag)       & $1.644\pm0.011$ & X, 10\\
    $W3$ (mag)        & $1.748\pm0.025$ & 5\\
    $W4$ (mag)        & $1.406\pm0.014$ & 5\\
    $P_{\rm rot}$ (days)   & $11.60\pm0.16$  & Y, 10\\
    Mass ($M_\odot$) & $0.80 \pm 0.02$ & 6 \\
    Radius ($R_{\odot}$) & $0.738 \pm 0.003$ & 7 \\
    $T_{\mathrm{eff}}$ (K) & $5062 \pm 29$ & 8 \\
    $\mathrm{[Fe/H]}$ (dex) & $-0.044\pm 0.058$ &  9\\
    Age (Gyr) & $1.1\pm0.1$ & 10\\
    \enddata
    \tablenotetext{a}{ICRS, epoch 2016.0.}
    \tablenotetext{b}{Proper motion in R.A. includes a factor of $\cos \delta$.}
    \tablerefs{(1)~\citet{gaia_collaboration_gaia_2023}; (2)~\citet{BailerJones2021}; (3)~\citet{Gray2006}; (4)~\citet{Mermilliod1997}; (5)~\citet{Cutri2013}.; (6)~\citet{gonzalez_parent_2010}; (7)~\citet{rains_precision_2020}; (8)~\citet{soubiran_gaia_2024}; (9)~\citet{rosenthal_california_2021}; (10)~This work.
    }
    \tablecomments{X = The 2MASS $JHKs$ photometry \citep{Skrutskie2006} for $\epsilon$ Eri is poor ($\sigma_{JHKs}$ $\simeq$ 0.3\,mag) and not provided. $JHK$ photometry from \citet[][SAAO system]{Carter1990}, \citet[][ESO system]{vanderBliek1996}, and \citet[][Johnson-Glass hybrid system]{Koornneef1983} was converted to 2MASS $JHKs$ via \citet{Carpenter2001} (ESO \& Johnson-Glass) and \citet{Koen2007} (SAAO). Values are means and rms values of the transformed $JHKs$ magnitudes. Y = adopted from mean of $P_{\rm rot}$ reported by \citet{Donahue1996}, \citet{Hempelmann2016}, \citet{Frohlich2007}, \citet{Fetherolf2023}, where the uncertainty was calculated as ($P_{\rm rot}$(max)$-P_{\rm rot}$(min)/2/$\sqrt{13}$), where the denominator is the number of independent seasonal rotation periods reported ($N = 14$) minus 1. 
    }
\end{deluxetable}

Now, 30+ years of precision radial velocity measurements combined with absolute astrometry show that \epseri hosts an $\approx$1~$M_{\rm Jup}$ planet at $\approx$3.5~au in a circular orbit nearly coplanar with the outer debris disk \citep{campbell_search_1988,hatzes_evidence_2000,benedict_extrasolar_2006,  mawet_deep_2019, llop-sayson_constraining_2021, thompson_revised_2025}. \epserib is, presently, the nearest confirmed gas giant exoplanet to the Solar System and an attractive target for direct imaging observations. It presents the tantalizing opportunity to investigate the atmosphere of a true Solar System gas giant-analog. Several attempts to image the planet across the $KLMN$ bands with Keck, MMT, VLT, and Spitzer have resulted in non-detections \citep{macintosh_deep_2003, heinze_deep_2008, janson_comprehensive_2008, janson_high-contrast_2015, mawet_deep_2019, llop-sayson_constraining_2021, pathak_high-contrast_2021}. Among the current generation of telescopes, JWST provides the most sensitive exoplanet search capability in the infrared wavelengths at the separation of \epserib. \citet{llop-sayson_searching_2025} conducted an imaging search with the JWST/NIRCam coronagraph in the F444W (3.881--4.982~$\mu$m) filter. While the observations achieved deep contrast performance at the expected separation of \epserib ($\approx4.8\times10^{-7}$ at 1\arcsec), the sensitivity at the planet's predicted position angle \citep[from the orbit in][]{thompson_revised_2025} was compromised by one of the bright lobes of the NIRCam PSF \citep{krist_jwstnircam_2009}.

In this paper, we present results from the most sensitive 4--5~$\mu$m imaging search for \epserib using the JWST/NIRCam coronagraph to date. For this program, the observations were specifically designed to avoid the bright lobes of the NIRCam PSF at the predicted planet position and achieve maximum contrast performance. The paper is organized as follows. Section~\ref{sec:sys-prop} presents an updated age estimate for for \epseri based on the latest gyrochronology relations. Section~\ref{sec:obs} summarizes the observational sequences executed with JWST, the data reduction, and contrast calibration. Section~\ref{sec:evo} derives the planet physical properties from evolutionary models and discusses possible atmosphere scenarios consistent with the data. Section~\ref{sec:ring} places constraints on a potential ring system around \epserib. Section~\ref{sec:future} presents future opportunities to directly image \epserib with the next generation of instruments. Finally, Section~\ref{sec:concl} lists our conclusions. The effect of systematic uncertainties on the age of \epseri\ and the choice of optimal reduction geometries for contrast calibration are discussed in Appendix~\ref{app:sys} and \ref{app:cc}, respectively.

\section{Stellar Rotation and Gyrochronological Age}
\label{sec:sys-prop}

Table~\ref{tab:prop} lists observational data and updated stellar parameters for \epseri from literature and in some instances recalculated. We re-evaluate the age of $\epsilon$ Eri considering its rotation in light of updated rotation/activity data for members of age-dated star clusters. 

The observation that low-mass stars in clusters of various ages show systematic signs of spinning down as they age has led to development of stellar rotation as an age indicator --- ``gyrochronology" \citep{barnes_ages_2007}. For K-type dwarf stars like $\epsilon$ Eri, it is one of the better calibrated and understood age indicators \citep{mamajek_improved_2008,soderblom_ages_2010,bouma_empirical_2023}. The bounty of rotation periods measured for low-mass members of age-dated star clusters, mainly from the Kepler and TESS missions, has led to refined calibrations of the empirical trends in rotation period as a function of age and color or effective temperature. \citet{bouma_empirical_2023} provide an updated calibration of gyrochronology using rotation data for 10 age-dated star clusters which they validate over the effective temperature range 3800\,K $<$ $T_{\rm eff}$ $<$ 6200\,K and ages of 0.08--4.0\,Gyr. These generously bound the previous $T_{\rm eff}$ and age estimates for $\epsilon$ Eri.

The rotation period of $\epsilon$ Eri has been measured by numerous studies in recent decades. Here we combine several modern measurements to estimate a median rotation period and uncertainty in order to calcuate a gyrochronological age for $\epsilon$ Eri. Differential rotation has been observed for $\epsilon$ Eri, with seasonal estimates ranging from $\sim$11 days to $\sim$12 days \citep{donahue_relationship_1996}. We combine multiple independent estimates from different methods and observatories, including a value using time-series photometry data from a single sector of TESS data \citep[][\Prot\, = $11.039\pm2.233$\,d]{Fetherolf2023} and 2 periods from MOST \citep[][\Prot\,= 11.35\,d, 11.55\,d]{croll_differential_2006}, and periods measured using Ca II H\&K measurements using 9 seasons of Mt.~Wilson Ca HK data \citep[][\Prot(min)\,=\,11.04\,d, \Prot(max)\,=\,12.18\,d]{donahue_relationship_1996} and a season of TIGRE data \citep[][\Prot\,=$11.76\pm0.08$\,d]{Hempelmann2016}. The average for these 14 seasons of rotation periods for $\epsilon$ Eri is \Prot\,=\,11.60\,d. As the measured rotation periods are heteroscedastic and vary by season via differential rotation, we adopt a conservative estimate of the uncertainty as $\sigma_{P_{\rm rot}}$ $\simeq$ $\{$\Prot(max) $-$ \Prot(min)$\}$/$\{2\cdot\sqrt{N-1}\}$ $\simeq$ 0.16\,d (1.4\% uncertainty).

\begin{figure}[!tb]
\centering
\includegraphics[width=0.95\linewidth]{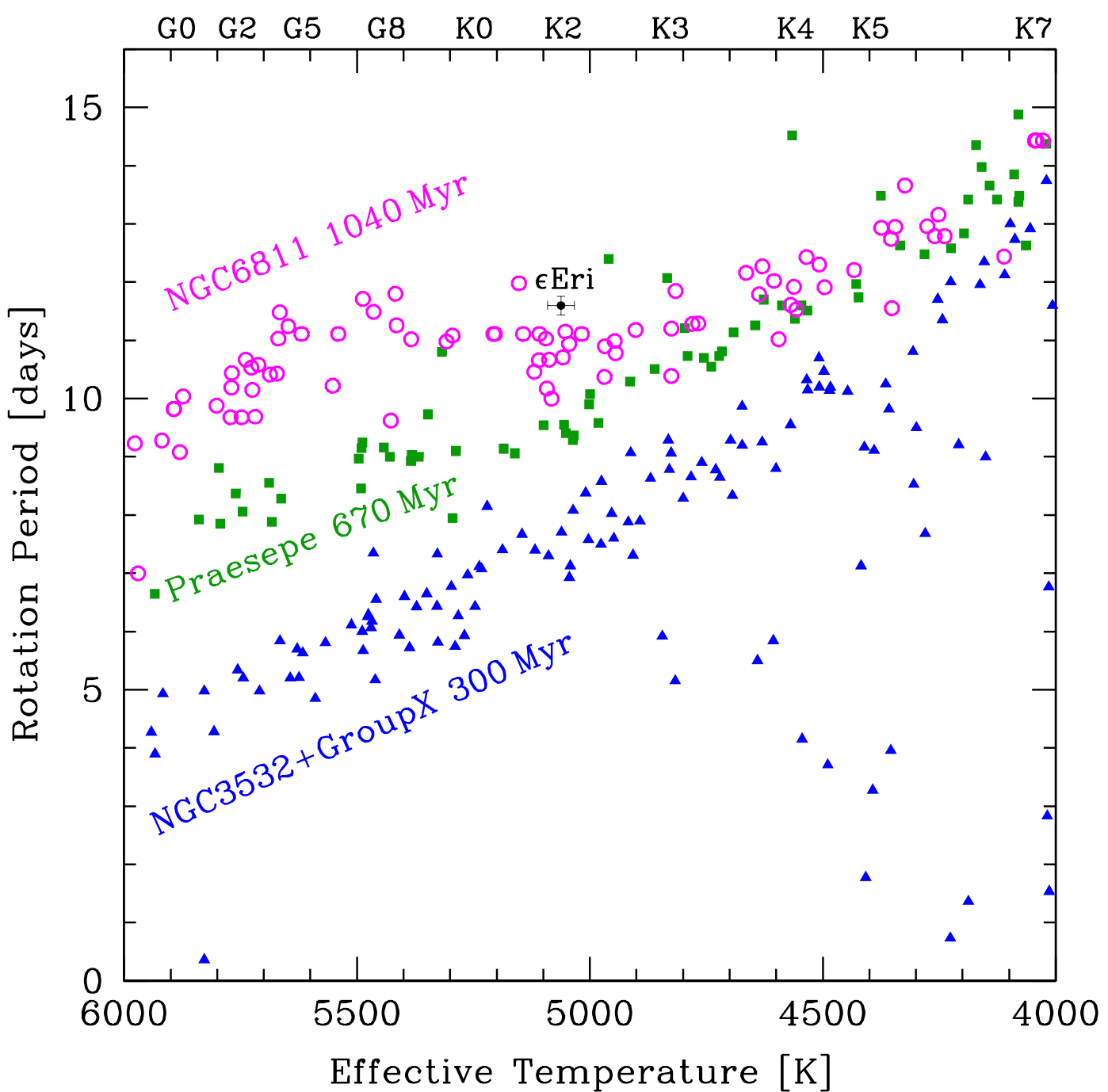}
\caption{Effective temperature (\Teff) versus rotation period for $\epsilon$ Eri compared to data from three young age-dated clusters compiled by \citet{bouma_empirical_2023}. 
Original sources of cluster data: 
NGC 6811 \citep{curtis_temporary_2019}, 
Praesepe \citep{rampalli_three_2021},
NGC 3532 \citep{fritzewski_rotation_2021},
Group-X \citep{messina_gyrochronological_2022}.
}
\label{fig:rotation}
\end{figure}

Using the gyrochronology age estimation code \texttt{gyro-interp} from \citet{bouma_empirical_2023}, and adopting our \Prot\, value and the 
effective temperature from \citet{soubiran_gaia_2024} (\Teff\, = $5062\pm29$\,K), we calculate a gyro-age of 1.119$^{+0.103}_{-0.096}$ Gyr ($\sim$9\%\, uncertainty). In Figure \ref{fig:rotation}, we overlay the $\epsilon$ Eri \Teff~vs.~\Prot\, data point with data for some star clusters with ages spanning 0.3--1 Gyr \citep[from Figure~1 of][]{bouma_empirical_2023}. One notes that $\epsilon$~Eri appears among the locus of points for the $1040\pm70$~Myr-old NGC 6811 cluster \citep{curtis_temporary_2019,bouma_empirical_2023}, where the early G to mid-K
dwarfs mostly have \Prot\, $\simeq$ 11--12\,d. This is the (\Teff, \Prot, age) regime where recent cluster studies have found that G and K dwarfs ``stall" in their spin down between ages of $\sim$0.7--1.4 Gyr \citep{curtis_when_2020}, and the stars spin down slower than predicted by older gyro-calibrations \citep[e.g.,][]{mamajek_improved_2008}. Ages less than 800 Myr, which bracket numerous adopted ages in the literature \citep[e.g.,][]{mamajek_improved_2008, coffaro_x-ray_2020}, are ruled out at $>$3$\sigma$ given this revised gyrochronology age (the impact of potential systematic uncertainties and the outlier possibility is discussed in Appendix~\ref{app:sys}). Based on this analysis, we adopt $1.1\pm0.1$ Gyr for the age of $\epsilon$ Eri. This is consistent with the $\sim$1~Gyr estimate derived by \citet{gai_modeling_2008} based on theoretical modeling of \epseri's fundamental parameters. The older age has significant implications for the effective temperature, and thus, expected F444W flux, of the planet. This is explored in more detail in \S\ref{sec:evo}.

A future avenue to obtain an independent age estimate to the one presented above is stellar asteroseismology. It is challenging to detect oscillations in main sequence stars cooler than the Sun given their small amplitudes, even in high precision space-based photometry \citep{kjeldsen_amplitude_2008, huber_testing_2011}, and separation from the instability strip. It is worth noting that a tentative detection of $p$-mode oscillations in \epseri was presented by \citet{noyes_evidence_1984} based on time variations (10~min period) in the star's Ca II H \& K emission lines. This has not been confirmed, but early modeling work suggested the observed frequency spacing ($\approx172\;\mu$Hz) could be reproduced using theoretical models for a stellar age of 1~Gyr \citep{soderblom_modeling_1989}, consistent with our gyro-based estimate. Radial velocities are much less affected by stellar granulation noise than photometry and potentially offer a more sensitive way to detect solar-like oscillations in cool K dwarfs. The predicted oscillation period for a K2 dwarf is $\approx4$~min \citep{christensen-dalsgaard_stellar_1983} and may be detectable with the current generation of high-resolution spectrographs (e.g., VLT/ESPRESSO, Keck Planet Finder, HARPS) at an exposure cadence of 45--60~seconds \citep[see example detections for $\alpha$~Cen~B, $\epsilon$~Ind~A, and $\sigma$~Draconis,][]{kjeldsen_solar-like_2005, campante_expanding_2024, lundkvist_low-amplitude_2024, hon_asteroseismology_2024}.

\section{Observations and Data Reduction} 
\label{sec:obs}
\subsection{Executed Sequences}
We obtained F210M (1.992--2.201~$\mu$m) and F444W (3.881--4.982~$\mu$m) JWST/NIRCam imaging of \epseri on UT 2025 August 29 as part of Cycle 4 program DD~9431 (PI: J.~Llop-Sayson, Co-PI(s): A.~Sanghi, W.~Thompson). The sequence consists of a two-roll angle observation centered on \epseri followed by imaging of reference star $\delta$~Eridani ($\delta$~Eri) with the \texttt{9-POINT-CIRCLE} dither pattern to improve sampling of the PSF \citep[see e.g.,][]{soummer_small-grid_2014, hinkley_jwst_2023}. This enables the use of both angular differential imaging \citep[ADI:][]{liu_substructure_2004, marois_angular_2006} and reference star differential imaging \citep[RDI:][]{lafreniere_hstnicmos_2009} to search for the giant planet \epserib. 

\begin{figure}
    \centering
    \includegraphics[width=\linewidth]{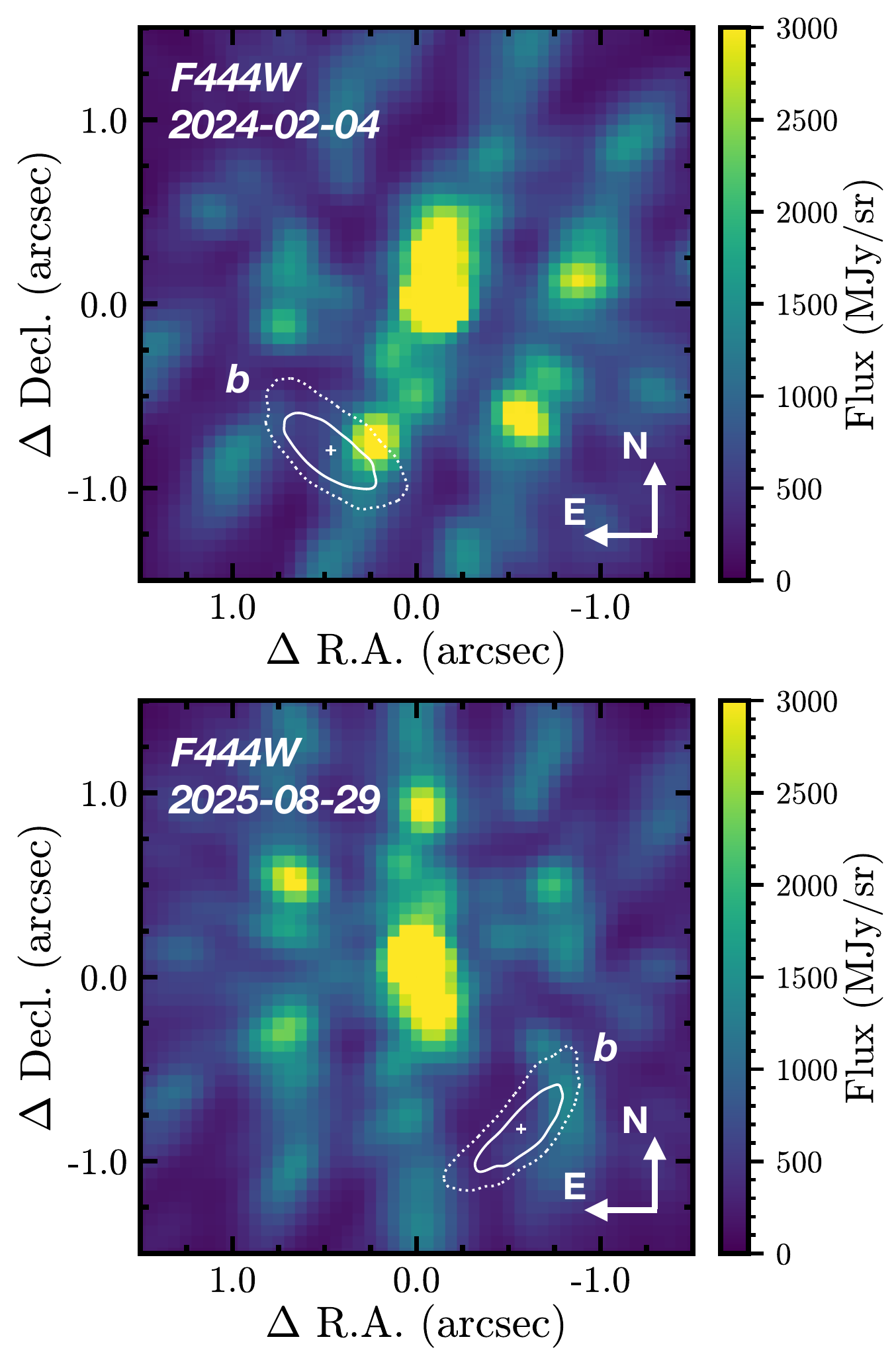}
    \caption{\emph{Top:} Raw coronagraphic image of \epseri from the observations presented in \citet{llop-sayson_searching_2025}. The solid and dotted white contours mark the 1$\sigma$ and 2$\sigma$ posterior predictions for \epserib's position, respectively. The expected position overlaps with one of the bright PSF lobes, which created strong residuals after PSF subtraction \citep[see Figure 3 in][]{llop-sayson_searching_2025}. \emph{Bottom:} Same as the top panel for the observations presented in this work. The expected position of \epserib\ does not overlap with the bright PSF lobes.}
    \label{fig:epoch-positions}
\end{figure}

The full set of observations were carried out between UT 2025 August 29 16:33:54 and 22:49:24 in a non-interruptible sequence for a total of $\approx$6.25 hours (including observatory overheads) simultaneously with the F210M/F444W filter combination and using the MASK335R coronagraph. Random centering errors in previous NIRCam observations, which observed \epseri at three roll angles but only used a \texttt{5-POINT-BOX} dither pattern for reference star observations \citep[][]{llop-sayson_searching_2025}, created strong (static) residuals at the location of the bright PSF lobes, overlapping with and degrading contrast performance at the expected planet position (Figure~\ref{fig:epoch-positions}). To avoid this issue and achieve the best contrast performance possible, the \epseri observations in this work were conducted at telescope V3 position angles 268.09$^\circ$ and 258.09$^\circ$ ($10^\circ$ roll), specifically chosen to ensure that the bright lobes of the NIRCam coronagraphic PSF do not coincide with the expected position of \epserib\ (Figure~\ref{fig:epoch-positions}).

We used the \texttt{BRIGHT2} readout pattern with 10 (10) groups per integration and 200 (40) integrations per exposure for \epseri ($\delta$~Eri). The total exposure time on \epseri is 142.5 minutes in F210M and F444W across the two-roll sequence. The total exposure time on $\delta$~Eri is 128.3 minutes in F210M and F444W across the nine point dither sequence.

\begin{figure*}
    \centering
    \includegraphics[width=\linewidth]{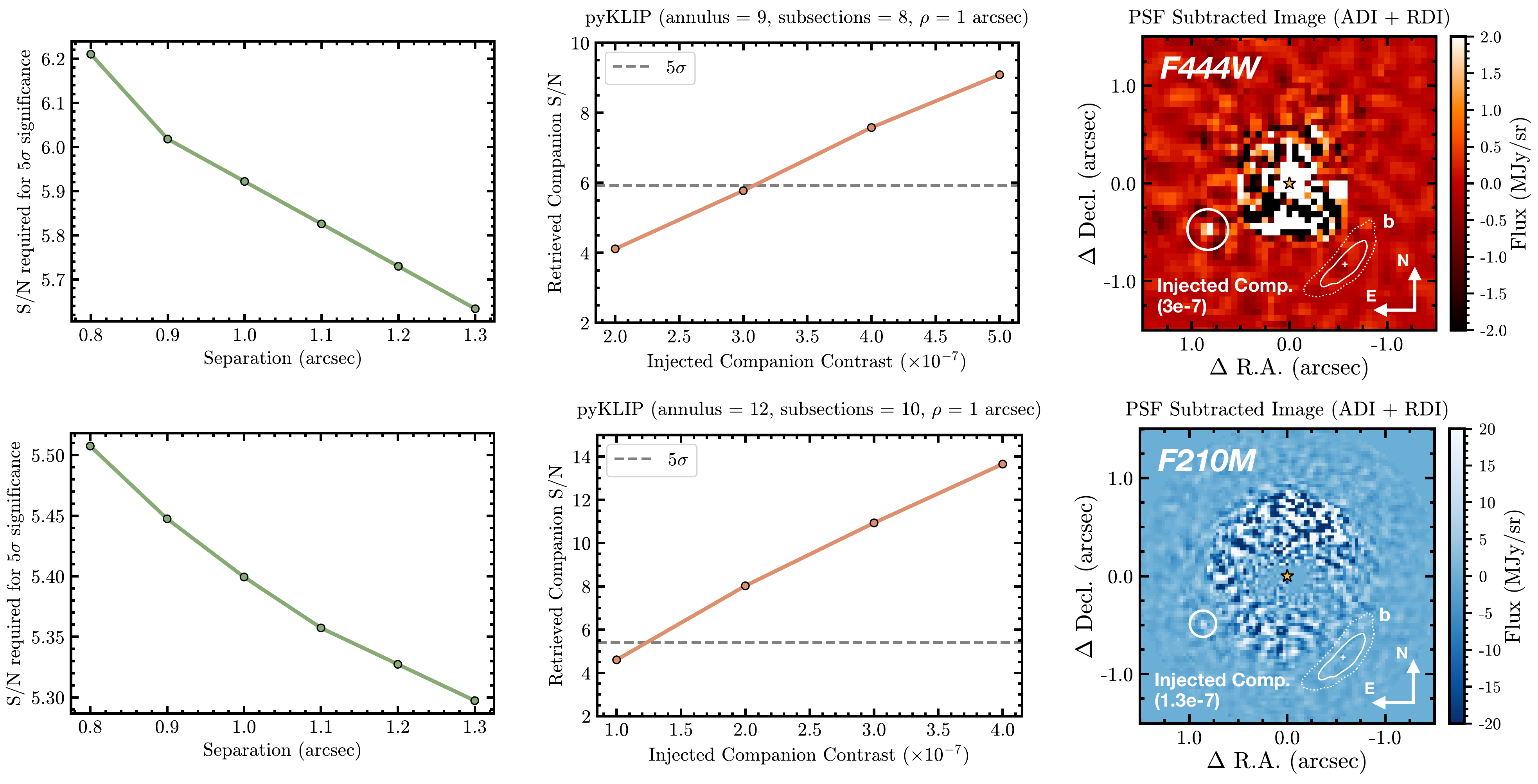}
    \caption{Upper limits on the F444W and F210M flux ratio of \epserib with respect to the host star from deep JWST/NIRCam imaging observations. \emph{Left column:} S/N measured in an image following \citet{mawet_fundamental_2014} that corresponds to a 5$\sigma$ significance detection as a function of separation for observations in F444W (top) and F210M (bottom). \emph{Center column:} Retrieved S/N for a synthetic companion injected at various contrast ratios at 1\arcsec\ separation, averaged across multiple position angles, for the reduction geometry noted in the title (using 10 principal components). The S/N equivalent to 5$\sigma$ significance is marked with a horizontal dashed line. \emph{Right column:} The ADI+RDI PSF subtracted images of \epseri in both filters for the reduction geometries in the center panel with an example synthetic companion injected at the contrast ratio corresponding to 5$\sigma$ detection significance. The expected location of \epserib\ on 2025 August 29 \citep{thompson_revised_2025} is shown with a `$+$' marker and the corresponding 1$\sigma$ (solid) and 2$\sigma$ (dotted) uncertainty contours. No significant point source signal is recovered at \epserib's expected location.}
    \label{fig:image-inj}
\end{figure*}

\subsection{Data Processing}
\label{sec:data-proc}
The uncalibrated Stage 0 (*uncal.fits) data products were acquired from the Barbara A. Mikulski Archive for Space Telescopes (MAST\footnote{The data described here may be obtained from the MAST archive at\dataset[10.17909/g6k6-bx64]{https://doi.org/10.17909/g6k6-bx64}.}) and processed with \texttt{spaceKLIP} \citep{kammerer_performance_2022, carter_jwst_2023, carter_spaceklip_2025}, a community developed pipeline for high contrast imaging with JWST. \texttt{spaceKLIP} wraps the \texttt{jwst} pipeline \citep{bushouse_jwst_2025} for basic data processing steps with modifications for coronagraphic imaging reduction. Data reduction via \texttt{spaceKLIP} for our observations implements best-practices discussed in various works \citep{kammerer_performance_2022, carter_jwst_2023, franson_jwstnircam_2024, gagliuffi_jwst_2025} and, more specifically, closely follows \citet{gagliuffi_jwst_2025}. 

We used \texttt{spaceKLIP} v2.1, via github commit \#11df3a1, and \texttt{jwst} v1.19.1; the calibration files were from CRDS v12.1.11 and the \texttt{jwst\_1413.pmap} CRDS context\footnote{\url{https://jwst-crds.stsci.edu/}}. The data were fit ``up the ramp" using the ``Likely" algorithm described in \citep{brandt_optimal_2024, brandt_likelihood-based_2024} and transformed from Stage 0 images into Stage 2 (*calints.fits) images, using a jump threshold of 4 and 4 pseudo-reference pixels on all sides of the subarrays. The $1/f$ noise was mitigated using a median filter computed along each column. We skipped dark current subtraction \citep{carter_jwst_2023}. Pixels flagged by the \texttt{jwst} pipeline (only the PSF core reached the saturation threshold) as well as 5$\sigma$ outliers detected using sigma clipping were replaced with a 2-dimensional interpolation based on a 9-pixel kernel. We also identified additional pixels affected by cosmic rays by flagging pixels with significant temporal flux variations across integrations and replaced them by their temporal median. Subsequently, the images were blurred above the Nyquist sampling criterion, using a Gaussian with a FWHM of 2.70 for the F444W/LW detector and 2.43 for the F210M/SW detector. These values ensure that sharp features in the data do not create artifacts when Fourier shifts are applied to the undersampled images. The position of the star behind the coronagraph was estimated by fitting a model coronagraphic point spread function (PSF) from \texttt{webbpsf\_ext} to the first science integration. All subsequent images were shifted by this initial offset. Images were cross-correlated to the first science integration, and these small shifts were applied to each integration to center the entire observing sequence to the first science integration. The position of \epseri behind the mask differed by $\approx$10~mas between the two rolls. We find that the alignment of the roll images with respect small-grid dither (SGD) pattern of the reference images is also favorable. The first roll PSF falls $\sim$10 mas away from the center of the SGD pattern, and the second roll only $\sim$6 mas from one of the offset reference images.

We carry out PSF subtraction with \texttt{spaceKLIP}, which uses the Karhunen-Lo\'eve Image Projection \citep[KLIP:][]{soummer_detection_2012} algorithm implemented in \texttt{pyKLIP} \citep{wang_pyklip_2015} to model and subtract the host star PSF. We search for \epserib using the ADI+RDI strategy with various combinations of the parameters \texttt{annulus} $=\{1, 2, 3, 4, 5, 6, 7, 8, 9, 10, 11, 12\}$, \texttt{subsections} $=\{1, 2, 3, 4, 5, 6, 7, 8, 9, 10, 11, 12\}$, and \texttt{numbasis} $=\{1, 2, 3, 4, 5, 6, 7, 8, 9, 10\}$. The central 1~FWHM region is masked and we set \texttt{annuli\_spacing} $=$ \texttt{linear} (annuli are distributed as a linear expansion in radial distance from the center) for all of the above reductions. The planet is not recovered in any of the above F210M and F444W reductions (representative examples in right panel of Figure~\ref{fig:image-inj}). Additionally, no other significant sources are detected in images.

\subsection{Contrast Calibration}
\label{sec:contrast}
We estimate the contrast sensitivity of our observations using injection-recovery tests, implemented with functionality in the \texttt{spaceKLIP} package, following the methods outlined in \citet{sanghi_worlds_2025}. PSF injections are carried out over the following separation ($\rho$) and position angle ($\theta$, East of North) grid: $\rho \in [0\farcs8, 1\farcs3]$ in steps of $0\farcs1$ and $\theta \in [0^\circ, 330^\circ]$ in steps of 30$^\circ$, excluding position angles overlapping with the expected planet position (210$^\circ$ and 240$^\circ$). The procedure is as follows.

\begin{enumerate}
    \item At a given $\rho$, synthetic companions are injected at varying $\theta$ and contrast ratios (between $7.5\times10^{-8}$ and $9\times10^{-7}$) in the raw coronagraphic image frames. The contrast is calculated with respect to a solar metallicity, $T_{\rm eff} = 5100$~K, log~$g = 4.50$~dex (cgs units) BT-NextGen stellar spectrum \citep{allard_model_2011, allard_models_2012} scaled by the interferometrically-determined radius of \epseri \citep[$0.738 \pm 0.003\: R_\odot$,][]{rains_precision_2020}. This model yields a stellar F444W flux of 40.7~Jy corresponding to a Vega apparent magnitude of 1.64~mag and a stellar F210M flux of 153.1~Jy corresponding to a Vega apparent magnitude of 1.63~mag.

    \item We perform ADI+RDI PSF subtraction for fixed \texttt{annulus} and \texttt{subsections} values, and \texttt{numbasis} $=\{1, 2, 3, 4, 5, 6, 7, 8, 9, 10\}$.

    \item We measure the signal-to-noise ratio (S/N) of the injected source using the method of \citet{mawet_fundamental_2014} and select the highest measured S/N across all \texttt{numbasis} reductions at each injection $\theta$.

    \item The S/N values are averaged across all $\theta$ values at a given $\rho$ to obtain the retrieved companion S/N as a function of injection contrast (center panel of Figure~\ref{fig:image-inj}).

    \item We determine the contrast level corresponding to companion detection at 5$\sigma$ significance. Note that this is distinct from the measured signal-to-noise ratio (S/N) of the injected source in the image. We use \texttt{vip\_hci}’s \texttt{significance} function \citep{gomez_gonzalez_vip_2017, christiaens_vip_2023} to calculate the S/N \citep[accounting for small sample statistics;][]{mawet_fundamental_2014} that corresponds to a 5$\sigma$ Gaussian detection significance (for the equivalent false positive probability, $2.9 \times 10^{-7}$) at each radial separation of injection (left panel of Figure~\ref{fig:image-inj}).
    
\end{enumerate}

\begin{figure}
    \centering
    \includegraphics[width=\linewidth]{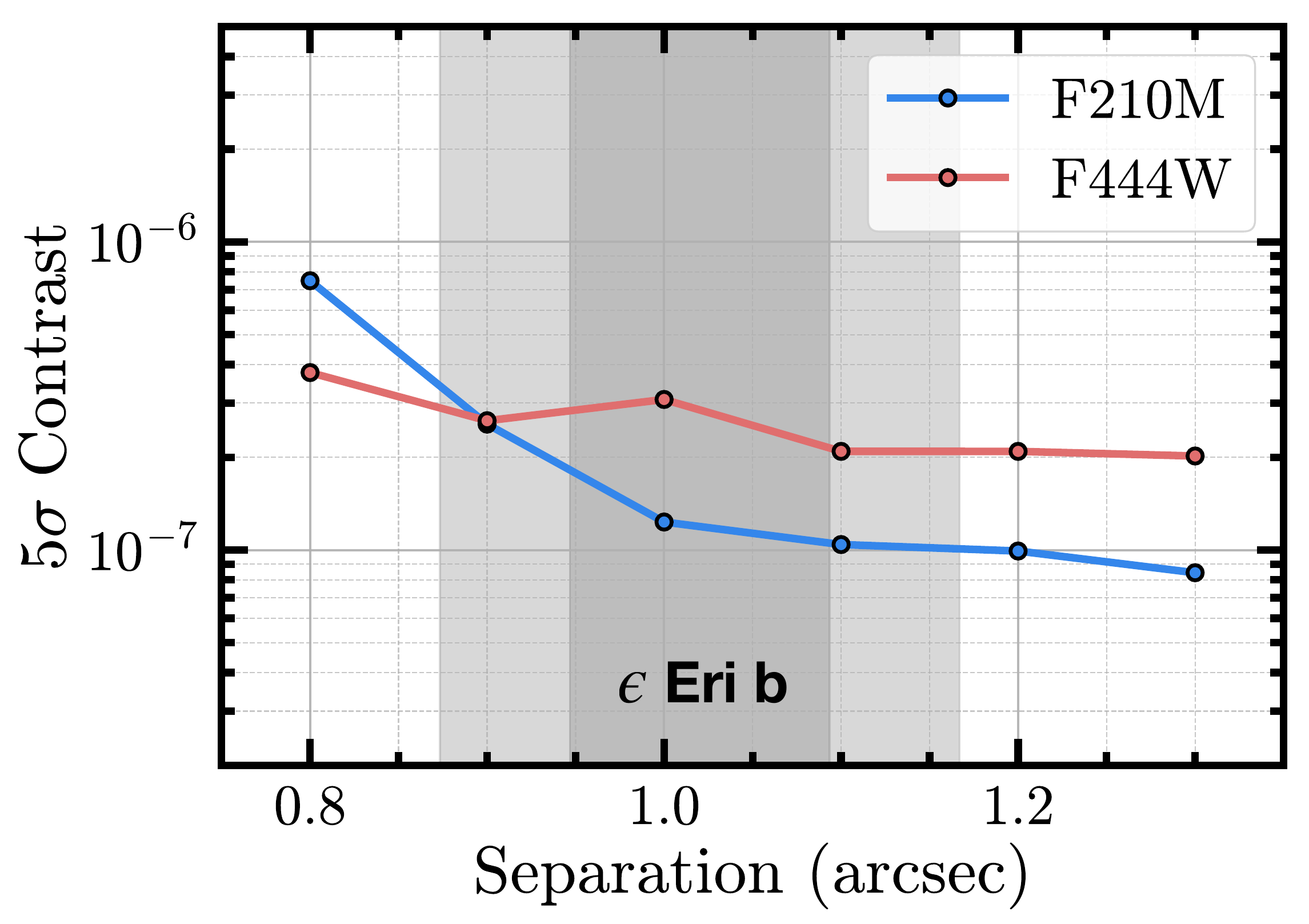}
    \caption{5$\sigma$ calibrated contrast curve for F210M (blue) and F444W (red) observations of \epseri. The dark and light gray shaded regions correspond to the 1$\sigma$ and 2$\sigma$ separation range expected for the planet.}
    \label{fig:image-cc}
\end{figure}

\noindent
The above procedure is repeated for each injection separation to estimate the 5$\sigma$ contrast curve for a given reduction geometry. The reduction geometries used are determined based on a preliminary sensitivity analysis discussed in Appendix~\ref{app:cc}. The 5$\sigma$ contrast curves from all reduction geometries are combined by selecting the best contrast performance at each separation to generate the final contrast curve shown in Figure~\ref{fig:image-cc}. Note that the above procedure to derive the optimal 5$\sigma$ contrast curve is different from the default method implemented in \texttt{spaceKLIP} (discussed in Appendix~\ref{app:cc}). The above strategy focuses on maximizing the detection S/N of an injected companion as opposed to minimizing the annular noise estimate.

The average 5$\sigma$ contrast sensitivity in the 1$\sigma$ separation range for \epserib is $\approx3.0\times10^{-7}$ ($\Delta = 16.3$~mag) in the F444W filter and $\approx1.3\times10^{-7}$ ($\Delta = 17.2$~mag) in the F210M filter (Figure~\ref{fig:image-cc}). The new F444W NIRCam imaging is $\approx$1.6$\times$ deeper at 1\arcsec\ separation compared to limits in \citet{llop-sayson_searching_2025}. However, when considering sensitivity to \epserib, the gain is significantly larger because our observations avoid the bright NIRCam PSF lobes (the contrast sensitivity reported in \citet{llop-sayson_searching_2025} does not apply to \epserib\ because of the presence of strong subtraction residuals at the expected position of the planet).

\begin{figure}
    \centering
    \includegraphics[width=\linewidth]{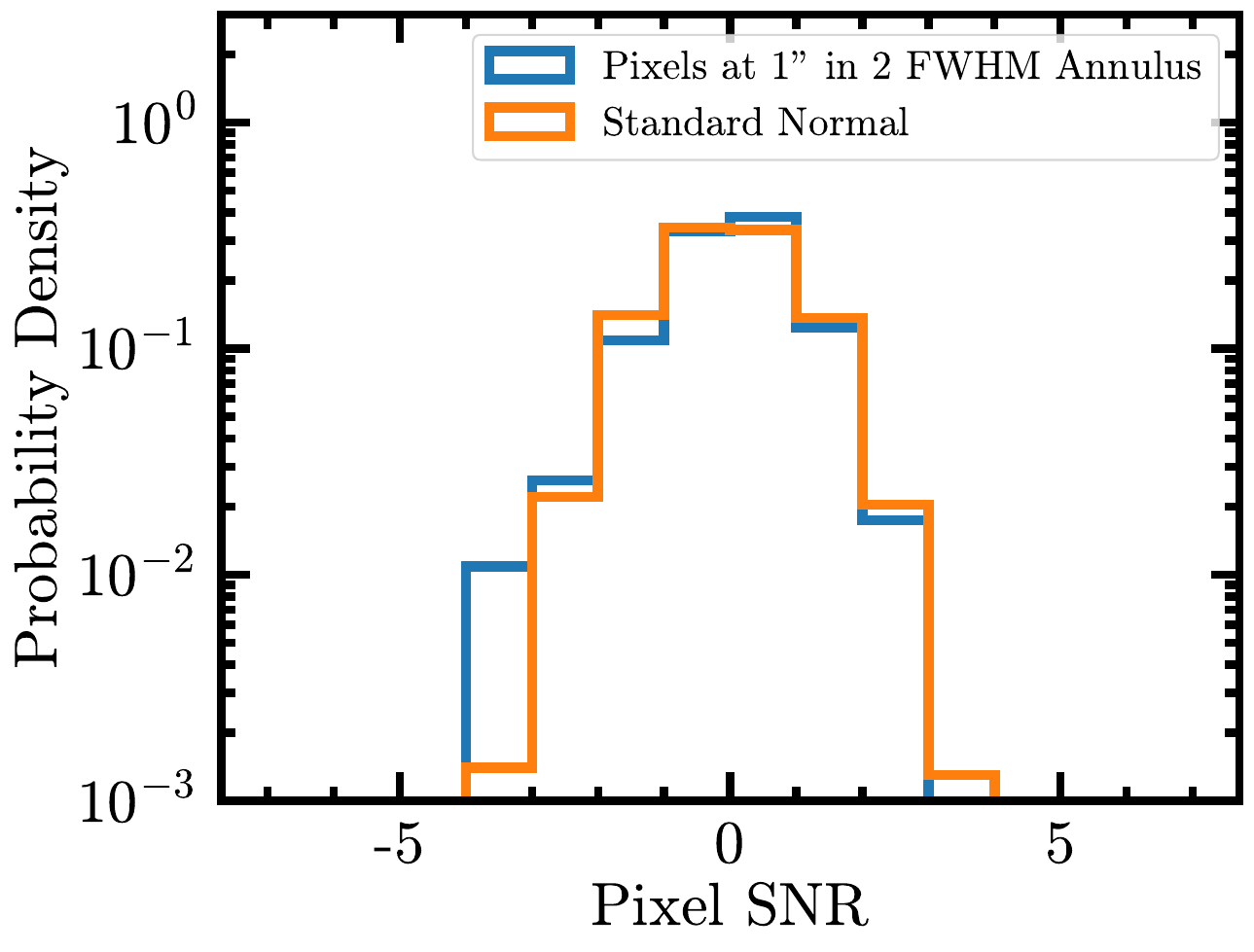}
    \caption{Distribution of signal-to-noise ratios (SNRs) of pixels from the F444W PSF-subtracted image in a 2~FWHM width annulus centered at a separation of 1\arcsec\ compared to a standard normal distribution.}
    \label{fig:snr}
\end{figure}

The F444W contrast curve in Figure~\ref{fig:image-cc} represents the deepest contrast performance achieved at $\sim$1\arcsec\ separations with NIRCam coronagraphy to date. A preliminary calculation suggests that the contrast performance is within a factor of six of the photon + read noise limit. It is important to note that, so far, the false positive rate (occurrence rate of non-astrophysical high signal-to-noise ratio features) in NIRCam coronagraphic images after PSF subtraction has not been characterized in literature. This can be done by comparing an ideal Gaussian probability density function (PDF) with the normalized histogram of pixel values in a signal-to-noise ratio (SNR) map calculated based on the post-processed image \citep[e.g.,][]{ruffio_improving_2017,thompson_deep_2023,  ruffio_jwst-tst_2024}. As an example, we calculate an SNR map \citep{mawet_fundamental_2014} for the post-processed F444W image of \epseri in Figure~\ref{fig:image-inj} and estimate the PDF of pixel SNRs in a 2~FWHM width annulus centered at 1\arcsec\ separation (a total of 460 pixel samples). The statistics of the noise is reasonably well-matched to a Standard Normal distribution, but does exhibit a higher occurrence in the left tail (Figure~\ref{fig:snr}). This excess of high SNR pixels indicates that a conventional ``5$\sigma$" detection threshold may not be equivalent to a false positive rate of $2.9 \times 10^{-7}$ (but a higher value). This is an important consideration when evaluating the significance of candidates identified after PSF subtraction. Here, and for all previously published results, it introduces some uncertainty in the reported 5$\sigma$ contrast curves. It is not possible to accurately estimate the false probability rate based on a single observation given the overall low chance of high SNR false probability events. Such a calculation with all available NIRCam data is beyond the scope of this work. Given this caveat, we proceed with interpreting the NIRCam 5$\sigma$ upper limit in the context of the planet's properties in the following sections. We do note that the well-constrained prediction for the planet's position based on the latest orbit solution \citep{thompson_revised_2025} helps reduce the impact of the unknown false positive rate on the calculated flux upper limit.

\section{Evolutionary and Atmospheric Model Analysis}
\label{sec:evo}
The estimated dynamical mass for \epserib from joint modeling of radial velocity and absolute astrometry data is $0.98 \pm 0.09\:M_{\rm Jup}$ \citep{thompson_revised_2025}. Combining this precise mass measurement with an age of $1.1\pm0.1$~Gyr, we first estimate the range of possible fundamental parameters for the planet using different evolutionary models. Next, given \epserib's fundamental parameters, the deep NIRCam upper limit on the planet's 4--5~$\mu$m flux enables us to constrain possible atmosphere scenarios for the planet. We investigate the consistency of both cloud-free and cloudy ($\rm{H_2O}$) atmospheres with the observations using the latest generation of low-temperature self-consistent grid models. Finally, we investigate the possibility of a lower planet mass, compared to the dynamical mass, as an alternate explanation for the NIRCam non-detection using flux predictions from the Sonora Flame Skimmer evolutionary models.

\begin{figure*}
    \centering
    \includegraphics[width=\linewidth]{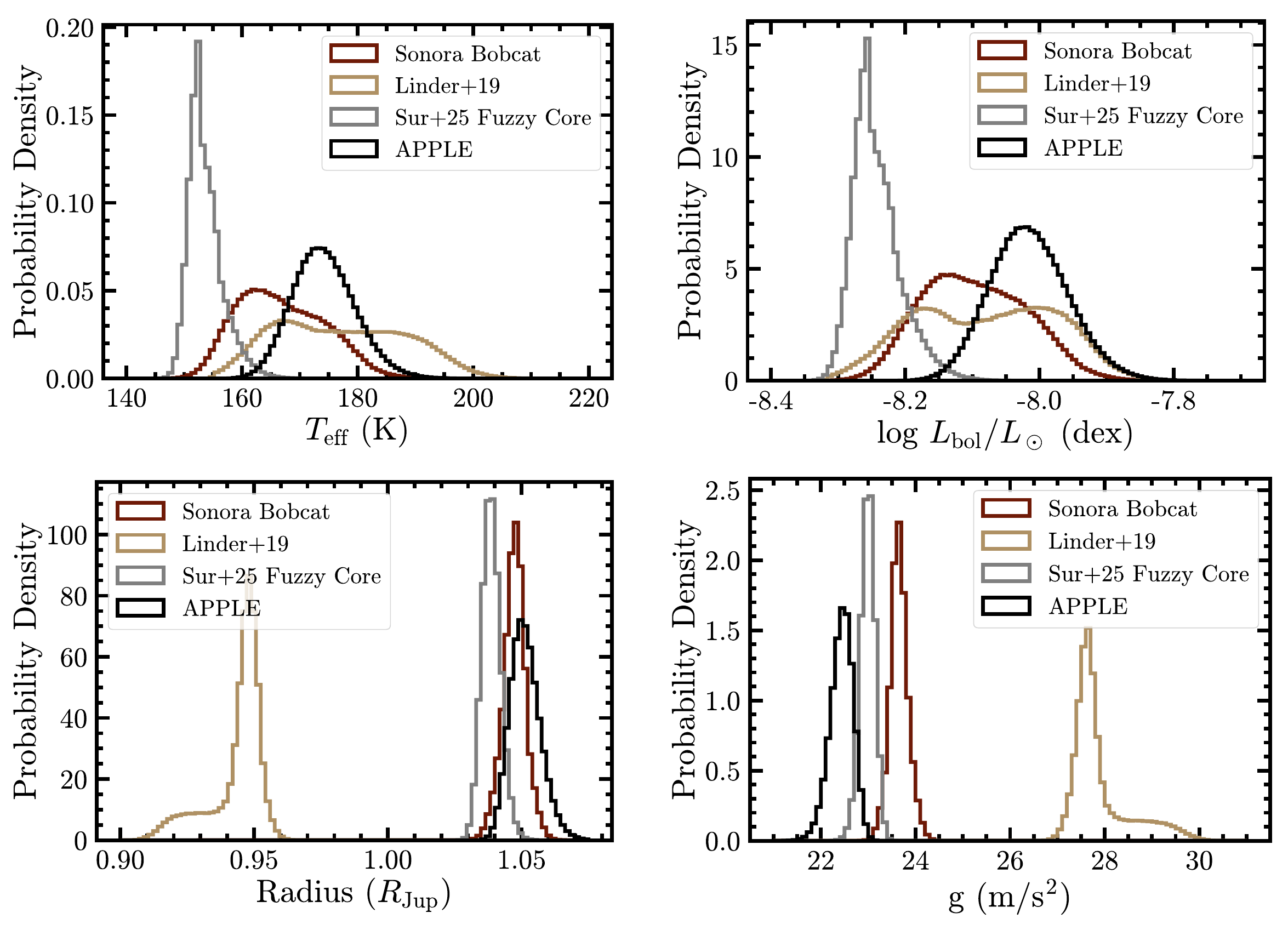}
    \caption{Inferred effective temperature ($T_{\rm eff}$), bolometric luminosity (log~$L_{\rm bol}/L_\odot$), radius, and surface gravity ($g$) for \epserib using the 1~$M_{\rm Jup}$ planet tracks from various evolutionary models.}
    \label{fig:teff-cooling}
\end{figure*}

\subsection{Fundamental Parameters for 1~$M_{\rm Jup}$ \epserib}
We compute evolutionary tracks for a 1~$M_{\rm Jup}$ planet using the planetary evolution code \texttt{APPLE} \citep{sur_apple_2024}, which incorporates the latest hydrogen–helium equation of state from \citet{Chabrier2021}, implemented via the publicly available module of \citet{Arevalo2024}. The models adopt updated atmospheric boundary conditions from \citet{Chen2023} for cloudless, non-irradiated atmospheres. The effective temperature is computed using the relation $T_{\rm eff}^4 = T_{\rm int}^4\,+\, T_{\rm eq}^4$, where $T_{\rm eq}$ is derived from the stellar bolometric luminosity evolution calculated using MIST stellar evolution tracks \citep{Paxton2011, Dotter2016, Choi2016} assuming a planetary orbital distance of 3.5~au. We adopt a Bond albedo of 0.5, consistent with the revised estimate for Jupiter by \citet{Li2018}. Two types of models are generated within this framework.

\begin{deluxetable}{llccccc}
    \tabletypesize{\footnotesize}
    \setlength{\tabcolsep}{2pt}
    \tablecaption{\label{tab:evo}Posterior 75\% Highest Density Interval}
    \tablehead{\colhead{Evolutionary Model} & \colhead{Model Ref.} &\colhead{$T_{\rm eff}$ (K)} & \colhead{Radius ($R_{\rm Jup}$)} & \colhead{$g$ ($\rm{m/s^2}$)} & \colhead{log~$L_{\rm bol}/L_\odot$ (dex)} & \colhead{$\rm{[M/H]}$\tablenotemark{a} (dex)}}
    \startdata
    Sonora Bobcat & \citet{marley_sonora_2021} & $[157, 175]$ & $[1.04, 1.05]$ & $[23.4, 23.9]$ & $[-8.19, -8.01]$ & $\{-0.5, 0.0, +0.5\}$\\
    Linder+19 (L19) & \citet{linder_evolutionary_2019} & $[163, 190]$ & $[0.94, 0.96]$ & $[27.2, 28.1]$ & $[-8.21, -7.95]$ & $\{-0.4, 0.0, +0.4, +0.8, +1.2\}$\\
    Sur+25 Fuzzy Core & \citet{sur_simultaneous_2025} & $[150, 156]$ & $[1.03, 1.04]$  & $[22.8, 23.2]$ & $[-8.29, -8.21]$ & \nodata \\
    \texttt{APPLE} & \citet{sur_apple_2024} & $[168, 180]$ & $[1.04, 1.06]$ & $[22.2, 22.7]$ & $[-8.09, -7.95]$ & $\{-0.5, 0.0, +0.5\}$\\
    \enddata
    \tablenotetext{a}{The metallicity range of tracks from each evolutionary model sets the minimum and maximum value of the uniform distribution used to sample metallicity. The exception is the Sur+25 Fuzzy Core model (single track, calibrated to Jupiter), which has the envelope metallicity evolve over time. Hence, no single value is quoted.}
\end{deluxetable}    

The homogeneous 1~$M_{\rm Jup}$ tracks computed with \texttt{APPLE} span three metallicities, with heavy elements uniformly mixed throughout the envelope at $Z$ = 0.00484, 0.0153, and 0.0484 (corresponding to [M/H] of $-0.5$, 0, and $+0.5$ dex, respectively), and are modeled using the AQUA equation of state \citep{Haldemann2020}. All homogeneous models are non-rotating, assume a compact core of 5~$M_{\oplus}$\footnote{This choice calibrates the homogeneous [M/H]$=+0.5$ model to reproduce Jupiter's radius at 4.56 Gyr.}, and an envelope helium abundance of $Y = 0.277$. 

Accurate gravity data from the Juno and Cassini missions and ring seismology for Saturn \citep{bolton_jupiters_2017,iess_measurement_2019,mankovich_diffuse_2021} show that the interior structure of both Jupiter and Saturn have inhomogeneous compositions with ``fuzzy cores" \citep[central regions enriched in heavy elements but not distinct from the deep interior, e.g.,][]{helled_fuzzy_2024}. Motivated by this, we include a more realistic Jupiter mass object with a fuzzy core  of mass 42~$M_{\oplus}$ that fits Jupiter's thermal evolution, atmospheric compositional constraints, and lower-order gravitational moments data \citep{sur_evolution_2025, sur_simultaneous_2025, Arevalo2025}. Unlike the homogeneous models, the fuzzy-core model does not assume a fixed envelope metallicity. Instead, its envelope composition evolves self-consistently with time due to convective mixing in the deep interior. These models represent significant improvements over the heritage Sonora Bobcat models as demonstrated in \citet{sur_next-generation_2025}. Although helium rain is implemented in \texttt{APPLE}, none of the models experience helium separation, as their present-day interior temperature profiles remain too high.

For completeness, we also consider the cloudless Sonora Bobcat \citep[][]{marley_sonora_2021} and cloudless \citet[][L19]{linder_evolutionary_2019} evolutionary tracks for a 1~$M_{\rm Jup}$ planet. The cloudless L19 models use \texttt{petitCODE} \citep{molliere_model_2015, molliere_observing_2017} to generate the atmosphere boundary conditions for planet cooling. We do not use the cloudy L19 tracks (note that these do not include the effect of water clouds) as they do not extend to the age range of \epseri. Sonora Bobcat tracks are available for planet metallicities $\left[\rm{M/H}\right] = \{-0.5, +0.0, +0.5\}$~dex and the L19 tracks are available for planet metallicities $\left[\rm{M/H}\right] = \{-0.4, +0.0, +0.4, +0.8, +1.2\}$~dex. 

\begin{figure*}
    \centering
    \includegraphics[width=\linewidth]{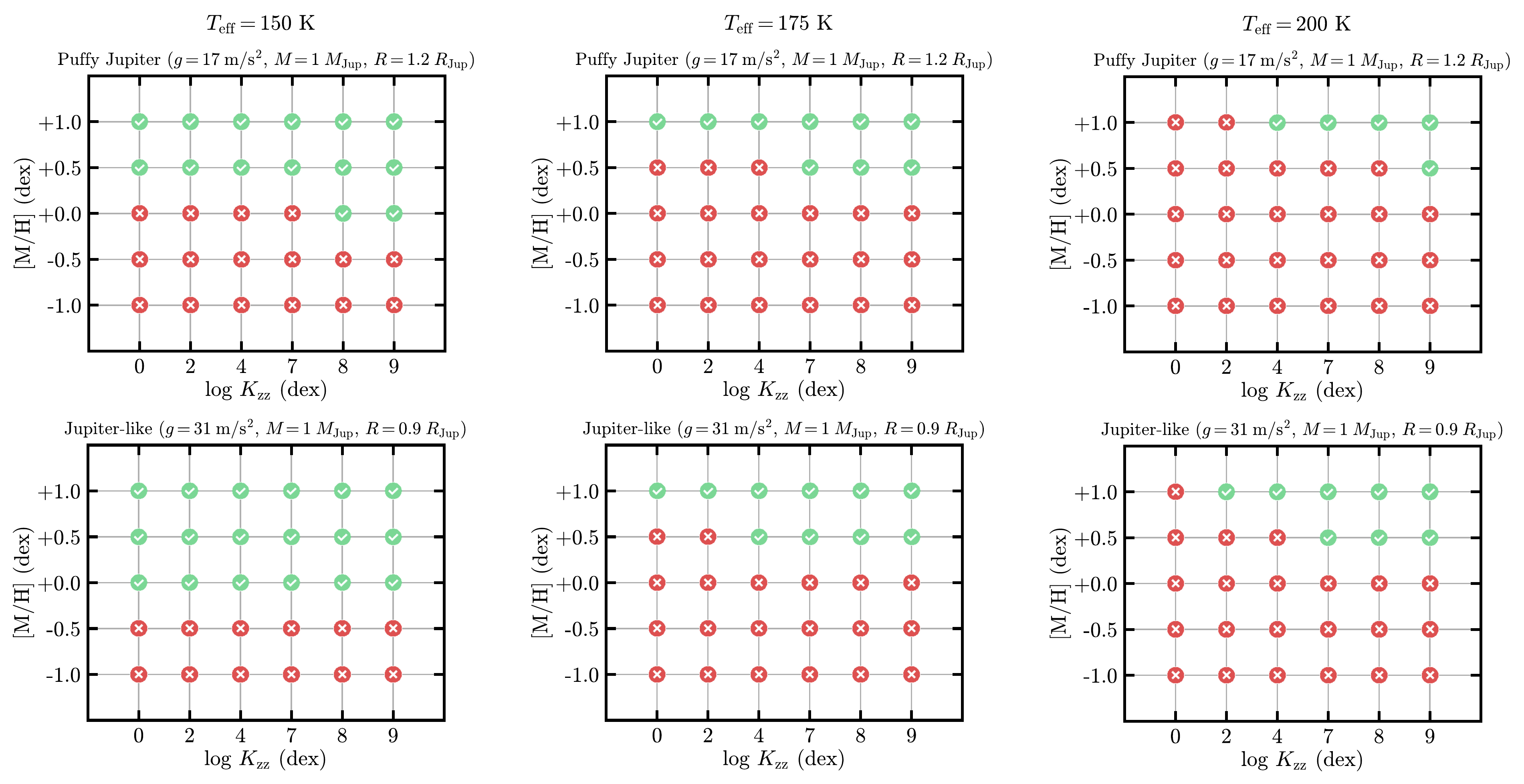}
    \caption{Consistency of Sonora Flame Skimmer cloud-free atmospheric models with NIRCam 5$\sigma$ upper limit. Each column considers models at a different temperature (150~K, 175~K, 200~K). The top row of models correspond to a puffy Jupiter scenario ($g = 17\;\rm{m/s^2}$) and the bottom row of models correspond to a Jupiter-like scenario ($g = 31\;\rm{m/s^2}$). A red `$\times$' indicates the model (with given $[$M/H$]$ and log~$K_{\rm zz}$, where we note that log~$K_{\rm zz}=0$~dex corresponds to a chemical equilibrium model for notational convenience) is inconsistent with (brighter than) the NIRCam 5$\sigma$ upper limit. A green `$\checkmark$' indicates the model is consistent with (fainter than) the NIRCam upper limit.}
    \label{fig:atmo}
\end{figure*}

Assuming a uniform distribution for planet metallicity across the range available for each evolutionary model and a normal distribution for the age, we linearly interpolate the 1~$M_{\rm Jup}$ tracks at $10^6$ (age, metallicity) samples to derive effective temperature, bolometric luminosity, radius, and surface gravity\footnote{For the fuzzy core model, the linear interpolation is only performed for age samples.}. Figure~\ref{fig:teff-cooling} presents the inferred parameter distributions. We quote the 75\% highest posterior density interval for each parameter in Table~\ref{tab:evo}. For any given evolutionary model, a lower (higher) temperature in the distribution corresponds to lower (higher) planet metallicity. We find that the inferred parameters are largely consistent between models. Notable differences include the prediction of a colder temperature by the fuzzy core model and a smaller radius ($<1\,R_{\rm Jup}$) by the L19 models. The fuzzy core represents a stably stratified region within the planet that inhibits convective heat transport from its interior to the atmosphere, resulting in a lower internal and effective temperature. In contrast, the homogeneous models, featuring a compact core and a uniformly mixed metal envelope, allow heat to be transported efficiently by convection. As a result, the homogeneous model produces a higher internal and effective temperature. The above assumes identical initial heat content/thermal energy and equilibrium temperatures.

An imaging/spectroscopic detection of the planet in the future with more sensitive high-contrast instruments can help further constrain the fundamental parameters and potentially test the fuzzy core hypothesis. The range of $T_{\rm eff}$ and radii predicted by the models are used to inform the atmosphere analysis in the next section. 

\begin{figure}
    \centering
\includegraphics[width=\linewidth]{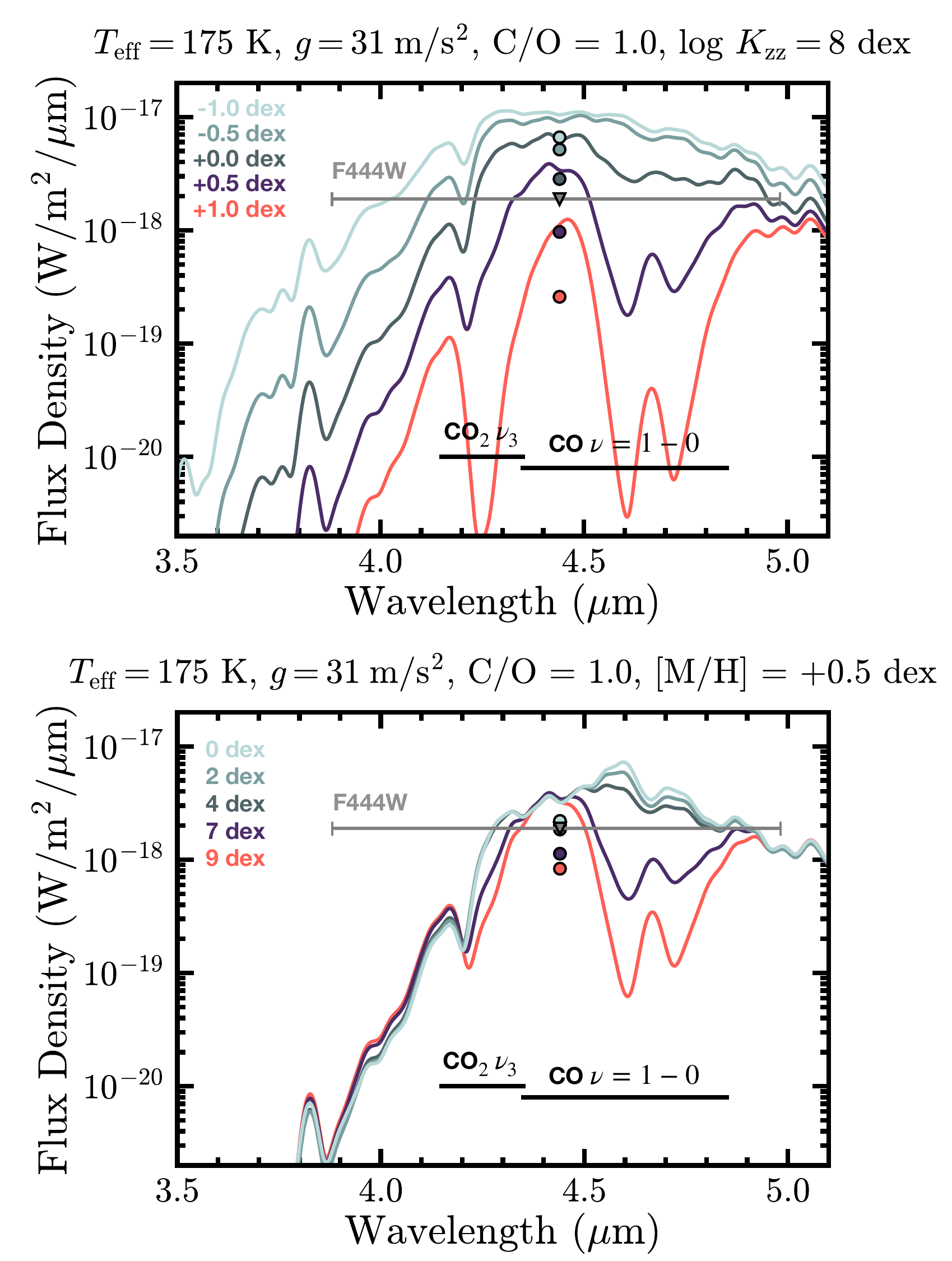}
    \caption{\emph{Top:} Sonora Flame Skimmer models for varying metallicity $[$M/H$]$ keeping other atmospheric parameters fixed. Circles show the synthetic F444W flux for each model. The inverted triangle marks the NIRCam 5$\sigma$ upper limit. Increasing metallicity boosts both the CO$_2$ and CO feature strengths and suppresses the F444W flux. \emph{Bottom:} Same as the top panel but for varying strengths of disequilibrium chemistry (log $K_{\rm zz}$), where 0~dex corresponds to a chemical equilibrium model. Increasing log $K_{\rm zz}$ (more vigorous mixing) boosts the strength of both the CO$_2$ and CO features and suppresses the F444W flux.}
    \label{fig:spec-cloudfree}
\end{figure}

\subsection{Cloud-free Model Grid}
\label{sec:cloud-free}
The Sonora Flame Skimmer models (Mang et al., in preparation) extend the cloud-free Sonora Elf Owl grid \citep{mukherjee_sonora_2024, wogan_sonora_2025} to colder effective temperatures, lower surface gravities, and a broader range of metallicities. These models include both equilibrium and disequilibrium chemistry and incorporate rainout chemistry for H$_2$O, CH$_4$, and NH$_3$—even in cloud-free atmospheres—similar to the treatment in Sonora Bobcat. They also address the underestimation of CO$_2$ found in the Sonora Elf Owl models \citep{mukherjee_sonora_2024}, which has since been revised in \citet{wogan_sonora_2025}. We use models spanning the $T_{\rm eff}$ range and bracketing the $g$ expected for \epserib from the evolutionary model analysis. The full set of parameter values are presented in Table~\ref{tab:models}. 

\begin{figure*}
    \centering
\includegraphics[width=\linewidth]{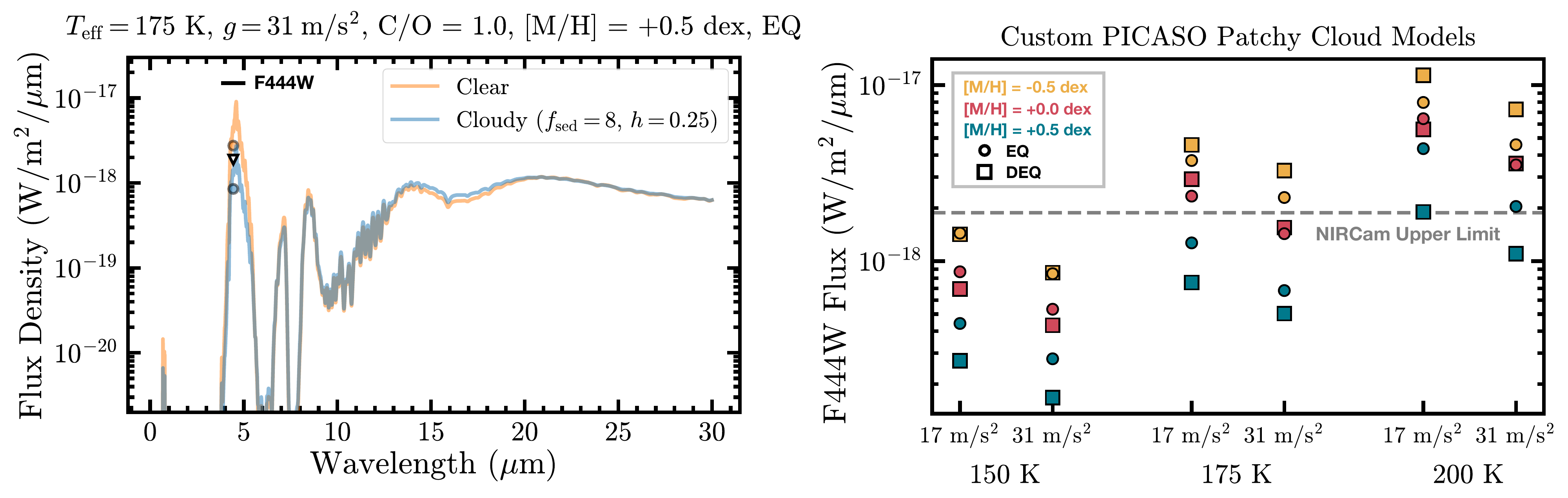}
    \caption{\emph{Left:} Comparison between an example Sonora Flame Skimmer model (orange) and a custom PICASO patchy water cloud model (blue) for identical atmospheric parameters (assuming a 1~$M_{\rm Jup}$ planet). Synthetic F444W photometry are shown as circles and the NIRCam F444W upper limit is shown as an inverted triangle. Water clouds suppress the 4--5~$\mu$m peak flux, redistributing it to longer wavelengths. \emph{Right:} Synthetic photometry for all patchy cloud models compared against the NIRCam upper limit.}
    \label{fig:patchy}
\end{figure*}

Our modeling approach is as follows. For a given surface gravity, we fix the radius of the planet in the model to the value that yields a planet mass of 1~$M_{\rm Jup}$, equal to the estimated dynamical mass \citep{thompson_revised_2025}. This corresponds to a radius of 1.2~$R_{\rm Jup}$ for $g = 17\; \rm{m/s^2}$, which we refer to as the ``Puffy Jupiter" scenario, and 0.9~$R_{\rm Jup}$ for $g = 31\; \rm{m/s^2}$, which we refer to as the ``Jupiter-like" scenario. Under these constraints, we synthesize the F444W photometry from the atmospheric model using the transmission curve from the Spanish Virtual Observatory (SVO) filter profile service\footnote{\url{https://svo2.cab.inta-csic.es/theory/fps/}}. If the synthetic photometry is brighter than the NIRCam 5$\sigma$ upper limit ($[$F444W$]$ $\approx$ 17.95 mag, Vega system), then we consider the model to be inconsistent with the non-detection. If the synthetic photometry is fainter than the NIRCam 5$\sigma$ upper limit, then we consider the model to be consistent with the non-detection. The consistency results for the complete set of models are summarized in Figure~\ref{fig:atmo}. We do not individually report consistency of models for varying C/O ratios as no trend was observed. Instead, if at least one model along the C/O axis, for given $[$M/H$]$ and log~$K_{\rm zz}$, is consistent with the upper limit, we mark it as such in Figure~\ref{fig:atmo}.

For a cloud-free atmosphere, the non-detection of \epserib is inconsistent with a sub-solar atmospheric metallicity and disfavors a solar atmospheric metallicity. As the temperature becomes warmer, the $[$M/H$]=$ +1.0~dex models are increasingly favored. The strength of the 4.3~$\mu$m $\rm{CO_2}$ feature strongly correlates with atmospheric metallicity \citep{lodders_atmospheric_2002, zahnle_atmospheric_2009, rustamkulov_early_2023, balmer_jwst-tst_2025}. A more metal-rich atmosphere is needed to account for the extent of flux suppression observed in F444W (Figure~\ref{fig:spec-cloudfree}). If we assume that \epserib has similar atmospheric metallicity as Jupiter ($[$M/H$]=$ +0.5~dex, 3$\times$ solar), then the cloud-free models additionally require strong disequilibrium chemistry (log~$K_{\rm zz} \geq 4$ dex), which increases the strength of both the 4.3~$\mu$m CO$_2$ and 4.5--4.8~$\mu$m CO absorption and contributes to the flux suppression (Figure~\ref{fig:spec-cloudfree}). The exception to the above is the 150~K effective temperature scenario (which is only likely for a fuzzy core model). In this case, a solar metallicity atmosphere can still explain the non-detection.

We note that the Sonora Flame Skimmer grid extends to $[$M/H$]=$ +1.5~dex (30$\times$ solar) and +2.0~dex (100$\times$ solar), and that our NIRCam upper limit is consistent with all models at these atmospheric metallicities. The star is approximately solar metallicity (Table~\ref{tab:prop}) indicating that significant atmospheric metal enrichment would likely originate from solid accretion during the planet formation process. Explaining the above values would thus require a large fraction of solids to be efficiently accreted and/or a high dust-to-gas ratio in the planet forming disk. We can also view \epserib as a younger analog of Jupiter and compare the two. Jupiter's atmosphere is enriched in metals only at 3$\times$ the solar value \citep{wong_updated_2004}. Even considering a lower mass planet, Saturn ($\approx3\times$ less massive than \epserib), for example, is enriched at $\approx$10$\times$ the solar level \citep[C/H,][]{fletcher_methane_2009}. For more massive planets, such as the four 6--9~$M_{\rm Jup}$ HR~8799 planets and 3--4~$M_{\rm Jup}$ AF~Lep~b, spectroscopic measurements find enrichment at levels $<10\times$ solar \citep[][]{balmer_vltigravity_2025, ruffio_jupiter-like_2026, xuan_compositions_2026}. Using the observationally-determined giant planet mass-metallicity relation from \citet{chachan_revising_2025}, we find that the \emph{bulk} metallicity ($Z_p$) of a 1~$M_{\rm Jup}$ planet is expected to lie between 2--10$\times$ the stellar value. The bulk metallicity of the planet is an upper limit on the atmospheric metallicity \citep[see e.g.,][]{thorngren_connecting_2019}. Current observational evidence thus suggests that significant atmospheric metal enrichment at 30--100$\times$ the solar value would be anomalous for \epserib.

\begin{deluxetable}{lc}
    \tabletypesize{\footnotesize}
    \tablecaption{\label{tab:models}Atmospheric Model Grids}
    \tablehead{\colhead{Parameter} & \colhead{Values}}
    \startdata
    \multicolumn{2}{c}{Sonora Flame Skimmer} \\
    \hline
    $T_{\rm eff}$ (K) & $\{150, 175, 200\}$ \\ 
    $g$ (m/$\rm{s^2}$)& \{17, 31\} \\ 
    $[\rm{M/H}]$ (dex) & $\{-1.0, -0.5, 0.0, +0.5, +1.0\}$ \\
    C/O ($\times$~solar) & $\{0.5, 1.0, 1.5, 2.5\}$ \\
    log $[$$K_{\rm{zz}}$ ($\rm cm^2\;s^{-1}$)$]$ & \{0, 2, 4, 7, 8, 9\} \\
    \hline
    \multicolumn{2}{c}{Custom PICASO Cloudy Models} \\
    \hline
    $T_{\rm eff}$ (K) & $\{150, 175, 200\}$ \\ 
    $g$ (m/$\rm{s^2}$)& \{17, 31\} \\ 
    $[\rm{M/H}]$ (dex) & $\{-0.5, 0.0, +0.5\}$ \\
    C/O ($\times$~solar) & $\{1.0\}$ \\
    log $[$$K_{\rm{zz}}$ ($\rm cm^2\;s^{-1}$)$]$ & \{0 (EQ), 7 (DEQ)\} \\
    $f_{\rm sed}$ & \{8\} \\
    $h$ & \{0.25\} \\
    \enddata
    \tablecomments{Chemical equilibrium models are noted above with log $[$$K_{\rm{zz}}$ ($\rm cm^2\;s^{-1}$)$] = 0$~dex for convenience only.}
\end{deluxetable}

\subsection{Cloudy Model Grid}
Water vapor is expected to condense into ice clouds in the atmospheres of the coolest extrasolar giant planets and brown dwarfs \citep[$\lesssim$400~K;][]{burrows_chemical_1999, marley_reflected_1999, burrows_beyond_2003, sudarsky_theoretical_2003, burrows_spectra_2004, sudarsky_phase_2005, morley_water_2014}. Indeed, water ice clouds feature prominently in Jupiter's atmosphere \citep{sato_jupiters_1979, carlson_abundance_1992, banfield_jupiters_1998}. In the context of NIRCam observations, the presence of water clouds can dramatically suppress the emergent 4--5~$\mu$m planet flux due to their increasing optical thickness with decreasing planet temperature \citep[e.g.,][]{morley_water_2014, mang_microphysics_2022, lacy_self-consistent_2023, mang_microphysical_2024, bowens-rubin_nircam_2025}. Given \epserib's temperature lies between 150--200~K, it is a strong candidate to host water ice clouds in its atmosphere. We test whether the NIRCam non-detection is consistent with this possibility.

We generate a custom grid of cloudy atmospheric models using \texttt{PICASO} \citep[Mang et al. in prep.]{batalha2019,mukherjee2023}, a 1D radiative–convective equilibrium climate model. \texttt{PICASO} now supports fully self-consistent cloudy atmosphere calculations through integration with \texttt{Virga} \citep{virga,virgapaper,virga2}, the Python implementation of the \citet{ackerman2001} cloud model. For these models, we adopt a parameterized patchy cloud framework \citep{marley2010,morley_spectral_2014, morley_water_2014}, in which the atmosphere is represented by two columns: a clear column and a cloudy column. Radiative–convective equilibrium is solved independently for each column, yielding emergent fluxes $F_{\rm clear}$ and $F_{\rm cloudy}$. The total emergent flux is then computed as a weighted sum of the two components, with the fractionation controlled by $h$, the fractional area of the clear atmosphere:

\begin{equation}
    F_{\rm total} = h F_{\rm clear} + (1 - h)F_{\rm cloudy}.
\end{equation}
Another relevant cloud parameter is the sedimentation efficiency, $f_{\rm sed}$, which sets the vertical extent of the water cloud. Generally, the optical thickness and vertical extent of the cloud decreases and particle size increases with increasing $f_{\rm sed}$.
For our grid, we use $h = 0.25$ and $f_{\rm sed} = 8$ for all models. For the considered planet temperatures (150--200~K), we encountered challenges with model convergence for smaller $h$ (greater cloud coverage fraction) and smaller $f_{\rm sed}$ (more optically thick) values. Previous work has shown that H$_2$O clouds can become extremely optically thick and introduce strong radiative feedback that complicates convergence in self-consistent atmospheric models \citep{morley_water_2014}. This motivated the use of a lower cloud fraction in the patchy cloud prescriptions to stabilize solutions. The full set of parameters explored in this mini-grid of cloudy models are summarized in Table~\ref{tab:models}. 

Following the modeling procedure described in \S\ref{sec:cloud-free}, we 
evaluated the consistency of the various patchy water cloud models with the NIRCam non-detection. The results are presented in Figure~\ref{fig:patchy}. The inclusion of water clouds suppresses the F444W flux in all models. For a subset, where the cloud-free counterparts were previously inconsistent (\S\ref{sec:cloud-free}), the water clouds bring the flux below the NIRCam upper limit (example in left panel of Figure~\ref{fig:patchy}). However, the suppression is not sufficient to make the 200~K, $< +0.5$~dex metallicity models consistent with the upper limit. For both $T_{\rm eff}=175$~K and $T_{\rm eff}=200$~K, the cloudy models still favor a metal-enriched atmosphere for \epserib. 

While the patchy cloud prescription represents a conservative scenario for cloud-induced flux suppression in the F444W bandpass, it provides a more realistic scenario for cold giant planets, as clouds are not expected to homogeneously blanket their atmospheres, analogous to the spatially heterogeneous band structures observed on Jupiter \citep{Carlson1992,Roos-Serote2000,Arregi2006}. Fully cloudy models would result in even greater flux suppression and would therefore remain consistent with the NIRCam upper limit.

In addition to H$_2$O, NH$_3$ can also condense to form clouds in the upper atmosphere of these cold Jupiter analogs \citep[e.g.,][]{burrows_nongray_1997, sudarsky_albedo_2000, sudarsky_theoretical_2003}. These clouds are expected to condense at low pressures, initially remaining optically thin and becoming thicker as atmospheric temperatures decrease \citep{morley_water_2014}. Although NH$_3$ clouds form well above the expected water cloud decks, 
H$_2$O clouds can strongly influence the atmospheric thermal structure (temperature profile of the upper atmosphere) and, in turn, affect whether NH$_3$ clouds are expected to form. NH$_3$ clouds could also condense onto H$_2$O ice particles, producing mixed cloud compositions \citep{hu_information_2019, guillot_storms_2020}. In this section, we presented preliminary results on the possible presence of water clouds in \epserib's atmosphere. Given the interplay of physical processes described above, modeling NH$_3$ clouds would be more relevant and informative once a direct detection of the planet is obtained.

\subsection{Lower Planet Mass as an Alternate Explanation for \epserib's Non-detection}

The goal of this section is to investigate what mass the NIRCam upper limit implies for \epserib if it's atmosphere is neither metal-enriched nor cloudy. Thus, for the following analysis, we use the solar metallicity, solar C/O, chemical equilibrium Sonora Flame Skimmer evolutionary models computed at three different masses (0.524~$M_{\rm Jup}$, 0.786~$M_{\rm Jup}$, and 1.048~$M_{\rm Jup}$). The Sonora Flame Skimmer evolutionary models use the previously described clear atmospheric models as surface boundary conditions and span masses from 15~$M_\oplus$ to 83~$M_{\rm Jup}$, with metallicities of [M/H] = {$-$1.0, $-$0.5, $+$0.0, $+$0.5, $+$1.0, $+$1.5, $+$2.0} dex. For objects with masses below 3~$M_{\rm Jup}$, the models include a 15~$M_\oplus$ core in the interior. These evolutionary models adopt the hydrogen and helium equations of state from \citet{chabrier_new_2019} and \citet{Chabrier2021}, and use the water equation of state from \citet{Mazevet2019} to account for all metals. The effects of these choices are discussed in \citet{chachan_revising_2025}. 

\begin{figure}
    \centering
\includegraphics[width=\linewidth]{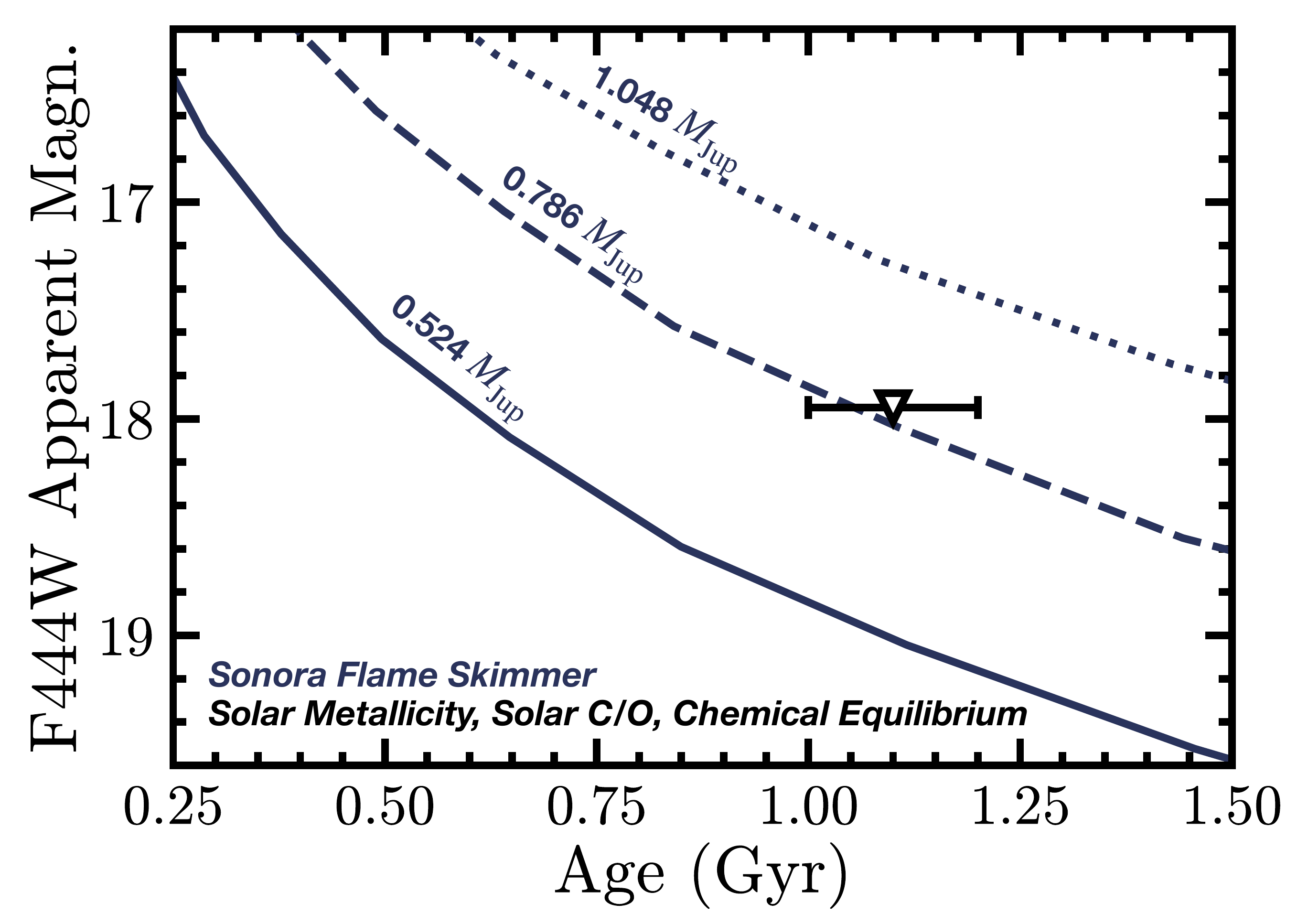}
    \caption{Predicted F444W magnitude using the clear Sonora Flame Skimmer evolutionary models for three different planet masses compared to the NIRCam upper limit (inverted triangle).}
    \label{fig:predicted-mag}
\end{figure}

We find that the 0.524~$M_{\rm Jup}$ and 0.786~$M_{\rm Jup}$ models are consistent and the 1.048~$M_{\rm Jup}$ model is inconsistent with the NIRCam upper limit (Figure~\ref{fig:predicted-mag}). Linearly interpolating planet mass as a function of F444W magnitude at ages sampled from a normal distribution ($1.1\pm 0.1$~Gyr), we find that \epserib's F444W magnitude would be equal to the measured upper limit for a mass of $0.81 \pm 0.05\;M_{\rm Jup}$. While masses below $0.81\;M_{\rm Jup}$ would be $>2\sigma$ away from the dynamical mass estimated by \citet{thompson_revised_2025} using their full data model, they would be consistent within 1$\sigma$ with the dynamical mass estimated using models excluding the HST/FGS and/or Gaia DR2 constraints. It is also worth noting that the evolutionary model-based mass limit agrees with the $0.65^{+0.10}_{-0.09}\;M_{\rm Jup}$ dynamical mass estimate by \citet{llop-sayson_constraining_2021} and the $0.76^{+0.14}_{-0.11}\;M_{\rm Jup}$ dynamical mass estimate by \citet{feng_revised_2023}. Given the challenging nature of fitting \epserib's orbit, we do not exclude the possibility that the planet's mass is $<1\;M_{\rm Jup}$ and that this may be a contributing factor in the NIRCam non-detection. Epoch astrometry from Gaia DR4 (expected December 2026) will help further refine the planet's dynamical mass to answer this question.

\section{Constraints on a Potential Planetary Ring System}
\label{sec:ring}
Ring systems are a characteristic property of the four giant planets in our Solar System. For mature exoplanets, rings can influence planetary bulk properties, atmospheric composition, and thermal structure \citep[e.g.,][]{odonoghue_ground-based_2016, waite_chemical_2018}. They may additionally hold clues to a giant planet's dynamical evolutionary history \citep[e.g.,][]{canup_origin_2010, hesselbrock_three_2019}. Recently, \citet{bowens-rubin_detection_2025} suggested the possibility of detecting large exoring systems in near-infrared (2~$\mu$m) reflected-light imaging with the NIRCam coronagraph. While \epserib was not detected in our F210M images, we can take advantage of the excellent contrast performance of the observations and knowledge of the planet's position to place the first constraints on the size of a potential ring system around the planet.

For the following analysis, we make similar assumptions as \citet{bowens-rubin_detection_2025}: (1)~the ring flux at $2\;\mu$m is dominated by starlight reflected by the ring with negligible contribution from its thermal emission; (2) the optical depth of the ring is high across the full area of the ring; (3)~thermal emission and reflected light from the planet are negligible compared to reflected starlight from the ring. Then, the flux ratio ($\epsilon$) of the planetary ring to the host star at phase angle $\alpha$ and wavelength $\lambda$ is given by,
\begin{equation}
\label{eq:contrast}
    \epsilon (\alpha, \lambda) = A_g (\lambda) g(\alpha, \lambda) \frac{\pi(R_{\rm out}^2 - R_{\rm in}^2)}{4\pi r^2},
\end{equation}
where $A_g(\lambda)$ is the geometric albedo of the ring, $g(\alpha, \lambda)$ is the phase function, $R_{\rm in}$/$R_{\rm out}$ are the inner/outer radii of the ring, and $r$ is the ring-host star separation. We note that $g(\alpha, \lambda)$ contains information about the geometry of the ring system with respect to our line of sight. Given that the planet is not detected in F210M, we do not attempt to constrain $g(\alpha, \lambda)$. Instead, we define the visible reflecting area $A_{\rm VR}$ as,
\begin{equation}
\label{eq:avr}
   A_{\rm VR} = g(\alpha, \lambda) \pi(R_{\rm out}^2 - R_{\rm in}^2).
\end{equation}
$A_{\rm VR}$ is thus a lower limit on the true surface area of a potential ring system. The maximum possible ring area is set by the planet's Hill radius. \epserib's Hill radius is $\approx$0.24~au \citep[Eq.~2 in][]{bowens-rubin_detection_2025} which corresponds to a maximum surface area $\approx9.2\times10^4$ times the area of Saturn's ring system. To calculate the area of Saturn's rings for comparison, we used $R_{\rm in,\:Sat} = 66,900$~km, corresponding to the inner edge of the D~ring \citep{murray_solar_1999, french_noncircular_2017}, and $R_{\rm out,\:Sat} = 136,780$~km, corresponding to the outer edge of the A~ring \citep{el_moutamid_how_2016}.

\begin{figure*}
    \centering
\includegraphics[width=\linewidth]{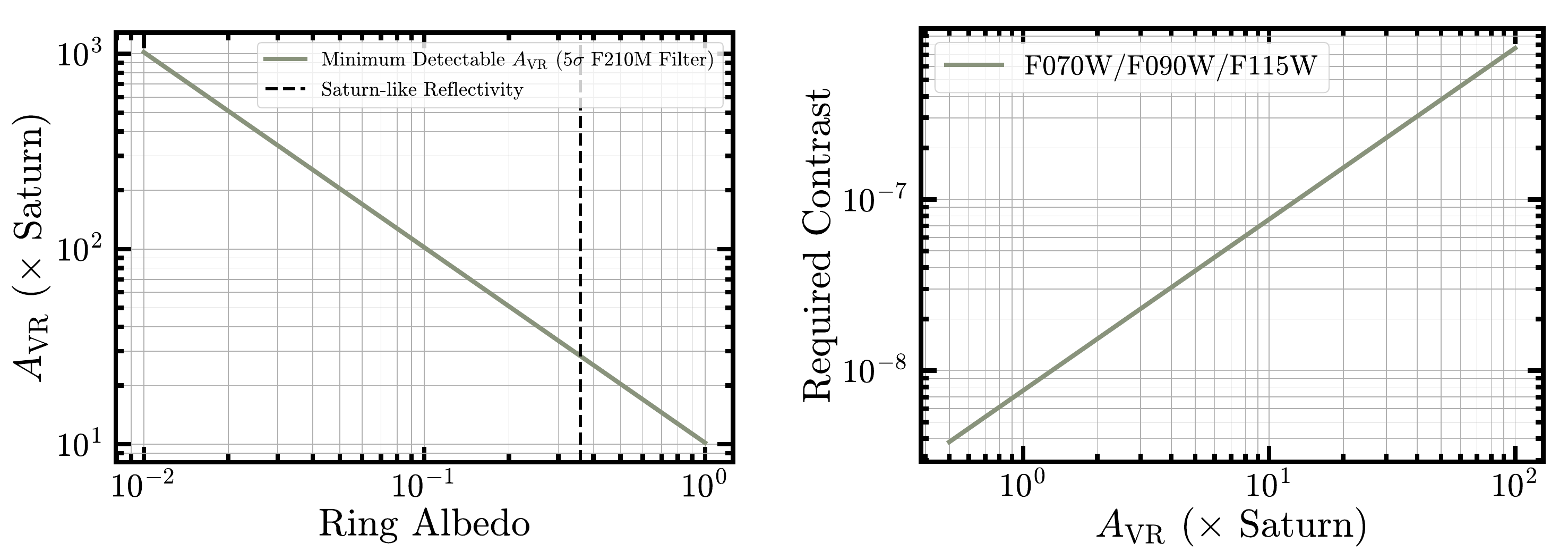}
    \caption{\emph{Left:} The minimum detectable visible reflecting area ($A_{\rm VR}$) as a function of albedo for a ring around \epserib based on the sensitivity in F210M filter. $A_{\rm VR}$ is expressed in terms of the area of Saturn's ring system. The reflectivity of Saturn's B~ring at $\approx$2~$\mu$m \citep{ciarniello_cassini-vims_2019} is marked with a dashed line for reference. \emph{Right:} Required contrast to detect exorings of various visible reflecting area in the F070W, F090W, and F115W filters \citep[albedo~$\approx 0.6$,][]{bowens-rubin_detection_2025}.}
    \label{fig:ring}
\end{figure*}

For a 5$\sigma$ contrast sensitivity of $1.3 \times 10^{-7}$ at 1\arcsec\ in F210M and planet-host star separation of $\approx$3.5~au, we can determine the minimum detectable $A_{\rm VR}$ for a ring around \epserib as a function of ring albedo using the above equations.  We find that our F210M observations rule out large exoring systems around \epserib, with $A_{\rm VR}$~$\gtrsim$~10--1000$\times$ the area of Saturn's rings depending on the albedo (sensitive to smaller ring systems for higher reflectivity; Figure~\ref{fig:ring}). For Saturn-like ring reflectivity, $A_{\rm VR} \lesssim$~30$\times$ the area of Saturn's rings. The above limits may be integrated as part of a larger sample of nearby systems observed with NIRCam at 2~$\mu$m to constrain the occurrence rate of large exoring systems.

We explore the sensitivity to exorings in the shorter wavelength NIRCam filters, specifically F070W, F090W, and F115W, under the same assumptions as the previous calculation. The albedo of Saturn's B ring is maximum in these filters, $\approx 0.6$ \citep[see Figure 2 in][]{bowens-rubin_detection_2025}. While these filters are unavailable for coronagraphy, potential observations may place the star at an off-detector position. Using Equations~\ref{eq:contrast} and \ref{eq:avr}, we can estimate the required contrast for detection as a function of the ring's visible reflecting area for $A_g = 0.6$. Achieving a contrast performance between $10^{-8}$--$10^{-7}$ (by taking advantage of the smaller PSF size and JWST's excellent stability) could enable the detection of exorings with visible reflecting area a few times the area of Saturn's ring (Figure~\ref{fig:ring}). 

\section{Future Opportunities for Direct Detection}
\label{sec:future}
The quest to image our closest Jupiter-analog, \epserib, is far from complete. With improved understanding of current instrument performance and systematics, the development of novel data processing techniques, and the upcoming first light of next-generation ground- and space-based facilities, numerous avenues are opening up to push current sensitivity limits and detect \epserib.

\begin{deluxetable*}{l|ccc|ccc}
    \tabletypesize{\footnotesize}
    \tablecaption{\label{tab:pred-flux}Predicted Contrast and Apparent Magnitude for \epserib in JWST and EELT/METIS Coronagraphic Imaging Filters}
    \tablehead{\colhead{Instrument/Filter} & \colhead{} & \colhead{Contrast} & \colhead{} & \colhead{} & \colhead{Apparent Magnitude (Vega System)} & \colhead{} \\ \colhead{} & \colhead{Worst} & \colhead{Mean} & \colhead{Best} & \colhead{Worst} & \colhead{Mean} & \colhead{Best}}
    \startdata
    NIRCam/F410M & $3.4\times 10^{-10}$& $9.9\times 10^{-9}$& $7.7\times 10^{-8}$ &25.2 & 21.6& 19.4 \\
    NIRCam/F430M & $3.1\times 10^{-9}$& $5.6\times 10^{-8}$& $3.2\times 10^{-7}$&22.9& 19.7 & 17.8 \\
    NIRCam/F460M & $6.8\times 10^{-10}$& $7.9\times 10^{-8}$& $9.3\times 10^{-7}$&24.6 & 19.4 & 16.7 \\
    NIRCam/F480M & $3.1\times 10^{-8}$& $2.0\times 10^{-7}$& $6.8\times 10^{-7}$&20.4& 18.4& 17.1 \\
    NIRCam/F444W & $1.2\times 10^{-8}$& $1.0\times 10^{-7}$& $2.9\times 10^{-7}$&21.4& 19.1&18.0  \\
    \hline
    MIRI/F1065C & $1.8\times 10^{-7}$& $5.1\times 10^{-7}$& $3.1\times 10^{-6}$&18.5& 17.3& 15.4 \\
    MIRI/F1140C & $5.1\times 10^{-7}$& $1.5\times 10^{-6}$& $1.1\times 10^{-5}$& 17.3& 16.1& 14.0 \\
    MIRI/F1550C & $2.7\times 10^{-6}$& $9.2\times 10^{-6}$& $5.2\times 10^{-5}$&15.5 & 14.2& 12.3 \\
    \hline
    METIS/$L$ & $1.2\times 10^{-13}$& $4.4\times 10^{-11}$& $6.5\times 10^{-9}$& 33.9& 27.5 & 22.1 \\
    METIS/$M$ & $1.1\times 10^{-8}$& $1.6\times 10^{-7}$& $7.1\times 10^{-7}$&21.6 & 18.6 & 17.0 \\
    METIS/$N2$ & $4.4\times 10^{-7}$& $1.2\times 10^{-6}$& $8.1\times 10^{-6}$& 17.5 & 16.4& 14.3 \\
    \enddata
    \tablecomments{All clear and cloudy models consistent with the NIRCam upper limit are considered together for the quoted contrasts and apparent magnitudes.}
\end{deluxetable*}

\subsection{Additional JWST Observations}
Can JWST still image \epserib? To address this question, we estimate the planet-to-star contrast ratio and \epserib's apparent magnitude in the NIRCam and MIRI coronagraphic imaging filters using all clear and cloudy models consistent with the F444W upper limit (\S\ref{sec:evo}). We report the worst case (most challenging), mean, and best case (most favorable) contrast and planet apparent magnitude in Table~\ref{tab:pred-flux} for each filter.

Among the NIRCam filters, F480M is the most promising choice for a future detection. The mean contrast of \epserib is $2.0\times10^{-7}$, just below the current F444W upper limit. In the best case scenario, the planet is $>2\times$ brighter than the sensitivity achieved here. However, we note that given the smaller filter bandwidth, F480M observations will require longer integration times than F444W observations to achieve the same detection sensitivity. One potential observation strategy is to conduct a multi-epoch search for \epserib. A multi-epoch approach would be particularly powerful for \epseri considering the extensive RV coverage and absolute astrometry data available for this system. For example, there are several ``de-orbiting" algorithms that can take advantage of the information content in a planet's orbit to recover signals that fall below the detection threshold in individual epoch images \citep{ruffio_improving_2017, nowak_k-stacker_2018,mawet_deep_2019,le_coroller_k-stacker_2020, le_coroller_efficiently_2022,thompson_deep_2023}. With these techniques, combining $N_{\rm obs}$ observations typically results in an SNR gain of $\sqrt{N_{\rm obs}}$, as expected for independent observations following Gaussian statistics. Note that since the contrast curves are a factor $\sim$6 above the photon + read noise limit, increasing the integration time will likely not boost contrast performance significantly. However, this may change given the progress in using differentiable optical models to accurately forward model JWST PSFs and achieve perfect subtraction down to the photon noise limit \citep[e.g.,][]{desdoigts_differentiable_2024,feng_exoplanet_2025}. Finally, recent work has demonstrated the capability of JWST/NIRSpec's Integral Field Unit (IFU) to spectroscopically detect faint companions at small angular separations \citep{ruffio_jwst-tst_2024, hoch_jwst-tst_2024, madurowicz_direct_2025, ruffio_jupiter-like_2026}. A Cycle~3 program (PID \#4982, PI: Ruffio) has observed \epseri using NIRSpec's IFU, targeting \epserib. This will be the most challenging detection attempted by NIRSpec so far. A sensitivity comparison with the NIRCam observations and identification of factors limiting contrast performance may also help identify future avenues for a 4--5~$\mu$m detection of \epserib. 

A MIRI detection of \epserib will be challenging. The typical contrast performance at 1\arcsec\ is $\gtrsim10^{-5}$ from previously published contrast curves in the three filters \citep{boccaletti_jwstmiri_2022, malin_first_2024, sanghi_worlds_2025, bendahan-west_jwstmiri_2025}, whereas the mean expected contrast is lower (Table~\ref{tab:pred-flux}). At $\sim$1\arcsec\ separations, the contrast performance is generally limited by PSF-subtraction residuals and increasing the integration time will likely not provide substantial gains. In the best case contrast scenario, a detection may be possible in F1550C. However, this does not consider contamination from \epseri's warm inner exozodi \citep{ertel_hosts_2020} and smooth outer dust disk, which was detected down to $\sim$3~au, with evidence for an additional disk component inside \epserib's orbit \citep{wolff_jwstmiri_2025}. The extended emission features would appear in the image after convolution with MIRI's coronagraphic PSF, making it challenging to disentangle a point source from the disk signal. There is ongoing work to leverage the MIRI/MRS IFU's spectroscopic capabilities to detect faint companions \citep[][]{patapis_direct_2022, malin_simulated_2023}, similar to advancements with NIRSpec. This approach potentially mitigates the issue of disk contamination given the planet and disk are spectrally distinct. Example practical challenges to overcome are saturation of the star on the detector and detector effects such as fringing.

\subsection{Roman Coronagraph Instrument}
The Nancy Grace Roman Space Telescope, planned to launch in late 2026, will be the first space-based observatory to demonstrate high-contrast imaging with active wavefront control using its Coronagraph Instrument \citep{kasdin_nancy_2020, poberezhskiy_roman_2021, mennesson_roman_2022}. Current best estimates of instrument performance on orbit achieve slightly better than $10^{-8}$ contrast \citep{cady_high-order_2025, zhou_high-order_2025}. This unlocks the exciting capability to directly image mature gas giants in reflected light for the first time. Here, we evaluate the opportunity to directly image \epserib with the Roman Coronagraph, in light of the JWST/NIRCam non-detection. We note that the following analysis is also broadly applicable to potential observations with the ExtraSolar Coronagraph (ESC) on the Lazuli Space Observatory \citep{roy_lazuli_2026}, a recently announced addition to the Eric and Wendy Schmidt Observatory System.

\subsubsection{Suitable Observing Modes}
The choice of imaging mode is driven by the planet-star projected separation (Figure~\ref{fig:phase}). Detection of \epserib requires the Shaped Pupil Coronagraph Wide Field-of-View (SPC WFOV) configuration (``best effort") to ensure the planet remains within the coronagraph's outer working angle \citep{riggs_flight_2025}. Among the two bandpasses available for SPC WFOV observations (Band~1, 575 nm; and Band~4, 825 nm; 10\% bandwidth\footnote{\url{https://roman.ipac.caltech.edu/docs/RomanCoronagraphPrimer_Current.pdf}}), the planet is always visible in Band~4 but only visible on certain dates in Band~1 (Figure~\ref{fig:phase}). Epoch astrometry from Gaia DR4 (expected December 2026) will help further refine the planet's orbit and reduce uncertainties on the projected separation before potential Roman observations.

\subsubsection{Predicted Planet-to-Star Flux Ratio}
\begin{figure}[!t]
    \centering
    \includegraphics[width=\linewidth]{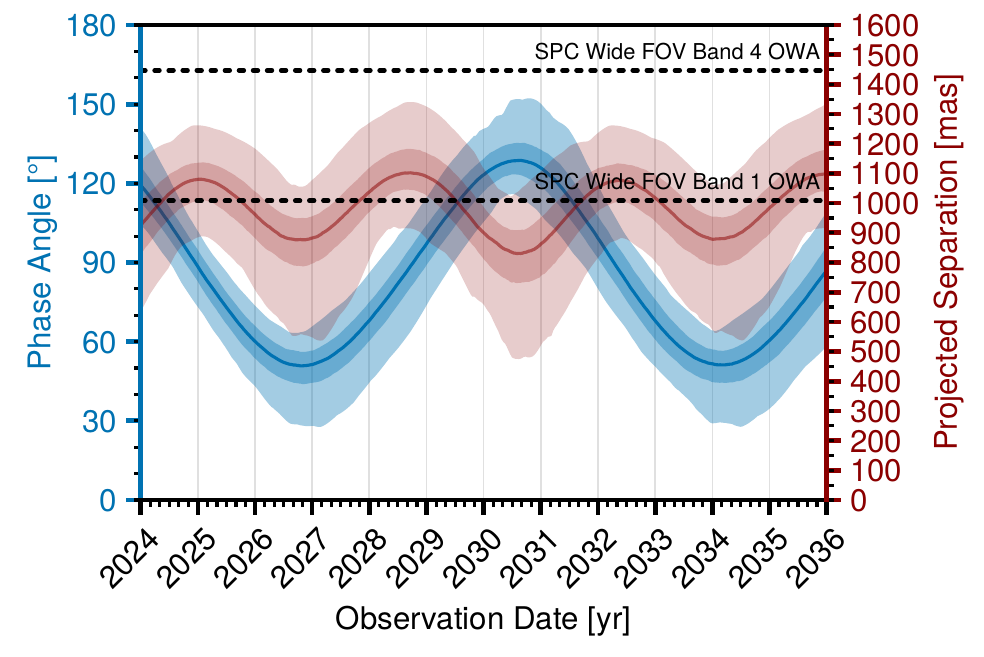}
    \caption{Phase angle (blue) and projected separation (red) of \epserib according to the latest orbital solution in \citet{thompson_revised_2025}. The two shades represent the 1$\sigma$ and 2$\sigma$ contours. Dashed lines mark the outer working angle (OWA) of the Roman Shaped Pupil Coronagraph (SPC) in Band~1 and Band~4 \citep{riggs_flight_2025}.}
    \label{fig:phase}
\end{figure}

The reflected light contrast depends on the planetary phase angle, $\alpha \in [0^\circ, 180^\circ]$, where full phase occurs at $0^\circ$. Using the precise orbital constraints from \citet{thompson_revised_2025}, we find that \epserib's phase angle oscillates between 50$^\circ$ and 130$^\circ$ (1$\sigma$, Figure~\ref{fig:phase}). We evaluate the contrast for all combinations of the above two phase angles, a clear and cloudy atmosphere, and the two planetary radii considered in previous sections. 

\begin{figure*}[!t]
    \centering
\includegraphics[width=\linewidth]{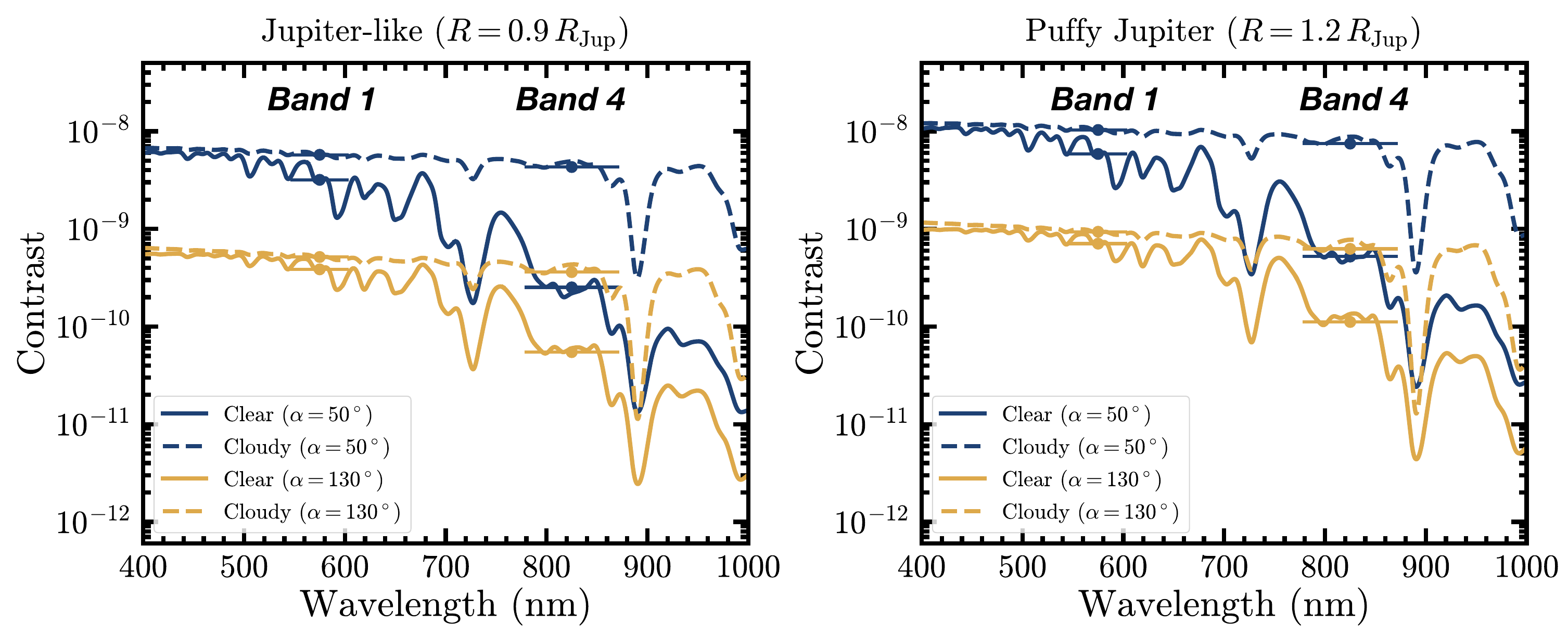}
    \caption{Reflected light spectra for \epserib computed with PICASO for possible planet radii, atmosphere types, and observable phase angles. The contrast in Roman bandpasses is most favorable for a cloudy planet at $\alpha = 50^\circ$.}
    \label{fig:contrast-reflected}
\end{figure*}

Reflected light spectra for \epserib covering the Roman Coronagraph bandpasses are generated using PICASO \citep{batalha_exoplanet_2019}. The input planet atmosphere (pressure-temperature profile, chemical composition, cloud properties) for the calculations use the $T_{\rm eff} = 150$~K, $g = \{17, 31\}\,\rm{m/s^2}$, C/O = 1$\times$ solar, [M/H]~$=+0.5$~dex, log~$K_{\rm zz} = 7$~dex Sonora Flame Skimmer clear profiles and custom PICASO cloudy ($h = 0.25$, $f_{\rm sed} = 8$) models, which are all consistent with the NIRCam upper limit for $M = 1\;M_{\rm Jup}$. As before, for the stellar spectrum, we assume using a solar metallicity, $T_{\rm eff} = 5100$~K, log~$g = 4.50$~dex (cgs units) BT-NextGen spectrum scaled by \epseri's radius. Table~\ref{tab:roman} summarizes the expected Band~1 and Band~4 contrast, and bond albedos, for the various planetary scenarios. Key takeaways are discussed below.

First, \epserib's phase angle is the most important parameter governing detectability. $\alpha = 50^\circ$ corresponds to the most favorable phase and $\alpha = 130^\circ$ corresponds to the least favorable phase in terms of contrast (Figure~\ref{fig:contrast-reflected}). Next, a cloudy atmosphere increases the planet's albedo and boosts the contrast in both bands. The difference is much more significant in Band~4. Finally, the contrast in Band~1 is not only more favorable than Band~4 but also less sensitive to the cloudiness of the atmosphere (Figure~\ref{fig:contrast-reflected}). Thus, Band~1 is the better choice for initial planet detection. Follow-up imaging in Band~4 could help distinguish between the cloudy and clear atmosphere scenarios.

\begin{deluxetable*}{cclllc}
    \tabletypesize{\footnotesize}
    \tablecaption{\label{tab:roman}\epserib Reflected Light Contrast and Albedo Predictions for Potential Roman Observations}
    \tablehead{\colhead{Planet Scenario} & \colhead{Phase Angle ($\alpha$)} & \colhead{Model Category} &\colhead{Band 1 Contrast} & \colhead{Band 4 Contrast} & \colhead{Bond Albedo}}
    \startdata
    & $50^\circ$ & Clear & $3.2 \times 10^{-9}$ & $2.5 \times 10^{-10}$ & 0.159 \\
    ``Jupiter-like" & & Cloudy & $5.7 \times 10^{-9}$ & $4.3 \times 10^{-9}$ & 0.341 \\
    (0.9~$R_{\rm Jup}$) & $130^\circ$ & Clear & $3.9\times10^{-10}$ & $5.5\times10^{-11}$ & 0.159 \\
    & & Cloudy & $5.2\times10^{-10}$ & $3.6\times 10^{-10}$& 0.341\\
    \hline
    & $50^\circ$ & Clear & $5.9 \times 10^{-9}$ & $5.2 \times 10^{-10}$ & 0.163\\
    ``Puffy Jupiter" & & Cloudy & $1.0 \times 10^{-8}$ & $7.5 \times 10^{-9}$ & 0.341 \\
    (1.2~$R_{\rm Jup}$) & $130^\circ$ & Clear & $7.1\times10^{-10}$ & $1.1\times10^{-10}$ & 0.163 \\
    & & Cloudy & $9.4\times10^{-10}$ & $6.2\times 10^{-10}$& 0.341 \\
    \enddata
\end{deluxetable*}

\subsubsection{Preferred Observing Window and Expected Exposure Time}
To maximize the likelihood of detection, \epserib should be observed in Band~1 as close to $\alpha=50^\circ$ as possible. An observation as early as the beginning of 2027, but before 2028, allows the planet to be imaged at near optimal phase while ensuring it remains within the OWA of the coronagraph in Band~1 (Figure~\ref{fig:phase}). The next available window for a similar configuration is the beginning of 2033. The target is, thus, appropriate for the ``observation phase", defined as an $\approx$18~month period after the initial 3~months of commissioning. 

The Roman Coronagraph exposure time calculator, \texttt{corgietc}\footnote{\url{https://github.com/roman-corgi/corgietc/tree/main}}, does not presently support calculations for Band~1 SPC WFOV observations. However, for reference, we note that detection at the 2$\times10^{-9}$ contrast level is possible with $<15$~hours of on-target integration time in the Band~4 SPC WFOV configuration based on \texttt{corgietc} calculations (assuming optimistic performance scenario).

\subsection{EELT/METIS}
The Mid-infrared Extremely Large Telescope Imager and Spectrograph \citep[METIS,][]{brandl_metis_2008, brandl_status_2018} will be one of the first-generation instruments on the 39~m European Extremely Large Telescope (EELT, first light $\sim$2029--2030). Operating at the diffraction limit, METIS will provide high-contrast imaging in the $LMN$ bands (3--13.5~$\mu$m) and high-resolution spectroscopy (R$\sim$100,000) in the $LM$ bands. In the background-limited regime, for a fixed SNR, the required integration time for detection scales as $t_{\rm int} \propto d^4/D^4$, where $d$ is the distance to the system and $D$ is the telescope diameter. Thus, a tremendous gain in sensitivity is expected for high-contrast imaging observations of the nearest stars with the $D=39$~m EELT in the mid-infrared. \epserib is a prime target for this instrument. \citet{bowens_exoplanets_2021} previously estimated that \epserib could be detected in only a few minutes of integration time with METIS. Given the revised older age and evidence of flux suppression from our NIRCam non-detection, we revisit the detectability of the planet in the ``HCI-L long" filter ($L$ band, 3.70--3.95~$\mu$m), the ``CO ref" filter ($M$ band, 4.70--4.90~$\mu$m), and the $N2$ filter (10.10--12.40~$\mu$m)

Using all clear and cloudy model atmospheres consistent with the NIRCam upper limit (256 total models), we calculate \epserib's apparent magnitude (Vega system) in all three filters, assuming constant filter transmission over the 50\% width \citep[][]{carlomagno_metis_2020}. The results are reported in Table~\ref{tab:pred-flux}. There is significant uncertainty in the planet flux. For an expected angular separation of $\approx$1\arcsec, \epserib is $\gtrsim10\;\lambda/D$ from the host star in the $L$, $M$, and $N2$ bands, and thus, in the background-limited regime for EELT/METIS observations \citep{carlomagno_metis_2020}. To estimate the integration time required to detect \epserib, we assume SNR $\propto f\cdot\sqrt{t_{\rm int}}$, where $f$ is planet flux, and a background-limited detection sensitivity (apparent magnitude) of 21.19~mag in the $L$ band, 18.44~mag in the $M$ band, and 15.14~mag in the $N2$ band for 1 hour of integration time \citep{bowens_exoplanets_2021}. The required integration times are then reported as a range for the minimum and maximum \epserib brightness and in parentheses for the mean \epserib brightness in each filter. We find $t_{\mathrm{int},\: M} \in [0.1, 337]$ (1.3)~hrs and $t_{\mathrm{int},\: N2} \in [0.2, 77]$ (10.2)~hrs. The planet is not detectable in $L$ band for any reasonable $t_{\rm int}$ given it's extremely cold temperature. 

$M$~band observations are the most efficient for detection at the average expected brightness but carry an overall larger uncertainty. This is because the degree of 4--5~$\mu$m flux suppression from enhanced metallicity, disequilibrium chemistry, and/or water ice clouds is currently unknown. The mid-infrared $N2$~band has a lower uncertainty, but the required $t_{\rm int}$ at average expected brightness is longer than possible in a single night's observation. An example observing strategy would be to first conduct reconnaissance imaging of the system in $M$~band. If the planet is not detected, a follow-up multi-night campaign in the $N2$~band can be conducted, similar to the NEAR experiment with VLT/VISIR \citep{wagner_imaging_2021, pathak_high-contrast_2021}. We note that it may also be possible to obtain a spectroscopic detection using cross-correlation techniques in the high-resolution $M$~band observation mode \citep[e.g.,][]{snellen_combining_2015}, but estimating an integration time requires more detailed simulations.

\section{Conclusions}
\label{sec:concl}
In this work, we presented the most sensitive search for the nearest Jupiter-analog exoplanet, \epserib, at 4--5~$\mu$m (F444W) using the JWST/NIRCam coronagraph. The planet continues to evade direct imaging detection, but, in the process, reveals more about itself. Our main conclusions are summarized below:
\begin{enumerate}
    \item The observations achieved a 5$\sigma$ contrast sensitivity $\approx3.0\times10^{-7}$ ($\Delta = 16.3$~mag) at the expected planet separation of $\approx$1\arcsec. This is the deepest contrast achieved with the NIRCam coronagraph at these separations to date.

    \item We combined multiple independent estimates in literature to calculate a rotation period of $11.60\pm 0.16\,\rm{d}$ for \epseri. Using the gyrochronology age estimation code \texttt{gyro-interp} from \citet{bouma_empirical_2023}, we updated \epseri's age to $1.1\pm0.1$~Gyr. This is significantly older than previous estimates between 400--800~Myr and impacts the expected temperature, and thus, flux, of the planet.

    \item For a dynamical mass $\approx1\,M_{\rm Jup}$ and the new gyro-age estimate, we found \epserib's $T_{\rm eff} \in [150, 200]$~K and $R \in [0.9, 1.1]\,R_{\rm Jup}$ using the latest generation of evolutionary models. Evolutionary model-predicted F444W fluxes ($\approx 17.3$~mag, assuming solar metallicity, solar C/O ratio, chemical equilibrium) for a 1~$M_{\rm Jup}$ planet are inconsistent with the NIRCam upper limit ($\approx 17.95$~mag) indicating that there is significant suppression of the 4--5~$\mu$m flux.

    \item A consistency analysis with cloud-free Sonora Flame Skimmer models and custom PICASO patchy water cloud models (bracketing the above fundamental parameters) found that the flux suppression can be explained by enhanced atmospheric metallicity and/or the presence of water ice clouds. This indicates \epserib's atmosphere shares similarities with Jupiter's atmosphere. Considering all models consistent with the upper limit, the F444W flux is expected to be between 18.0--21.4~mag, with a mean value of 19.1~mag.  

    \item An alternate explanation is that the planet is $<1\;M_{\rm Jup}$. Using the Sonora Flame Skimmer solar metallicity evolutionary models, we found that \epserib would have to be $\lesssim0.81\;M_{\rm Jup}$ to explain the NIRCam non-detection. This limit is 2$\sigma$ away from the latest dynamical mass ($0.98 \pm 0.09\;M_{\rm Jup}$) but consistent with other estimates in literature.
    
    \item The NIRCam/F210M observations achieved a 5$\sigma$ contrast sensitivity $\approx1.3\times10^{-7}$ at \epserib's separation. Assuming an optically thick ring with negligible 2~$\mu$m thermal emission and Saturn-like ring albedo ($A_g = 0.36$), the F210M non-detection ruled out ring systems with visible reflecting areas $\gtrsim$30$\times$ the area of Saturn's rings. We also found that achieving a contrast limit of $10^{-8}$--$10^{-7}$ in the shorter wavelength NIRCam filters (F070W, F090W, F115W) could enable the detection of exorings with visible reflecting area a few times the area of Saturn's ring.

    \item There is an opportunity at the beginning of 2027, but before the beginning of 2028, to image \epserib with the Roman Coronagraph near its most optimal phase angle ($\alpha = 50^\circ$), requiring a Band 1 contrast between $10^{-9}$--$10^{-8}$ for detection. This range encompasses both clear and cloudy atmosphere scenarios and the various possible planet radii. Follow-up imaging in Band 4 after an initial detection could help constrain the planet's cloudiness. We also discussed additional opportunities with JWST and EELT/METIS.
\end{enumerate}

This JWST/NIRCam search represents yet another step down the contrast ladder and towards the direct detection and characterization of mature gas giants. As a younger analog to Jupiter, \epserib remains one of the most exciting targets for detailed atmospheric characterization, investigating giant planet formation, studying planet-disk interactions, and conducting comparative planetology studies to place the Solar System in the broader exoplanetary context. 

\begin{acknowledgments}
The author thanks William Balmer and Rodrigo Ferrer-Chavez for helpful discussions on JWST/NIRCam data processing, Jerry Xuan and Heather Knutson for feedback on the atmospheric analysis, Jason Wang for suggestions on the evolutionary model analysis, Dan Huber for an initial assessment of periodicity in \epseri's TESS photometry, and Luke Bouma for helpful discussions on gyrochronology. We thank the referee for a prompt report and helpful comments that improved this manuscript. We gratefully acknowledge the support of the STScI Director's Office for this Director's Discretionary Time Program and the planning and operations team for the timely scheduling of this program. This work benefited from discussions during ExSoCal 2025 hosted at UCLA. This material is based upon work supported by the National Science Foundation Graduate Research Fellowship under Grant No.~2139433. Part of this research was carried out at the Jet Propulsion Laboratory, California Institute of Technology, under a contract with the National Aeronautics and Space Administration (80NM0018D0004).
\end{acknowledgments}

\begin{contribution}
A.~Sanghi led the overall analysis and the writing and submission of this manuscript. J.~Llop-Sayson designed the observation sequences and conducted an independent reduction of data. E.~Mamajek performed the age analysis for \epseri. J.~Mang generated the atmospheric models used in this work. W.~Thompson calculated the predicted planet positions and phases used in this work. A.~Sur generated the \texttt{APPLE} evolutionary models used in this work. All authors assisted with the preparation of the original JWST proposal and provided feedback on the manuscript.
\end{contribution}

\facilities{JWST(NIRCam)}

\software{\texttt{astropy
} \citep{astropy_collaboration_astropy_2013, astropy_collaboration_astropy_2018, astropy_collaboration_astropy_2022}, \texttt{matplotlib
} \citep{hunter_matplotlib_2007}, \texttt{numpy
} \citep{harris_array_2020}, \texttt{pandas
} \citep{mckinney_data_2010, team_pandas-devpandas_2025}, \texttt{python
} \citep{van_rossum_python_2009}, \texttt{scipy
} \citep{virtanen_scipy_2020, gommers_scipyscipy_2023}, \texttt{astroquery
} \citep{ginsburg_astroquery_2019, ginsburg_astropyastroquery_2024}, \texttt{scikit-image
} \citep{van_der_walt_scikit-image_2014}, \texttt{STPSF
} \citep{perrin_simulating_2012, perrin_updated_2014}, \texttt{jwst} \citep{bushouse_jwst_2025}, \texttt{pyKLIP} \citep{wang_pyklip_2015}, \texttt{vip} \citep{gomez_gonzalez_vip_2017, christiaens_vip_2023}, \texttt{spaceKLIP} \citep{kammerer_performance_2022, carter_jwst_2023, carter_spaceklip_2025}, \texttt{gyro-interp} \citep{bouma_gyrointerp_2023}, and \texttt{webbpsf\_ext} \citep{leisenring_webbpsf_2025}.}

\appendix
\restartappendixnumbering

\section{Effect of Systematic Uncertainties on Age and Effective Temperature}
\label{app:sys}
In \S\ref{sec:sys-prop}, we obtained an age estimate of $1.1 \pm 0.1$~Gyr from stellar gyrochronology. Ages $<800$~Myr are ruled out at $>3\sigma$ for this revised age. However, there are two sources of uncertainty that may reduce the precision of the above age estimate. First, the analysis used a rotation period with a 1.4\% uncertainty and a stellar effective temperature with a 0.6\% uncertainty. Potential systematic errors in the above input parameters to \texttt{gyro-interp} can modify the age posterior. For example, if we assume a 5\% uncertainty on the rotation period (due to biases induced by variation in spot latitudes) and 2\% uncertainty on the stellar effective temperature \citep[comparable to the systematic noise floor estimated by][]{tayar_guide_2022}, the estimated age is $1.10^{+0.14}_{-0.17}$~Gyr. This brings a younger age (700--800~Myr) within the 3$\sigma$ range. Additionally, in Figure~\ref{fig:rotation}, we observe a few outlier cases in Praesepe, stars with slower rotation periods than \epseri. This raises the question whether \epseri\ could similarly be a younger but slowly rotating outlier. The current model for intrinsic dispersion in the rotation period-effective temperature relations does not include the Praesepe outliers \citep{bouma_empirical_2023}. However, the astrophysical reason for the existence of these outliers is not known and it is not possible to confidently ascertain whether \epseri\ belongs to this category of objects. As such, based on the above, if we assume a younger age for \epseri (e.g., 700--800~Myr), the inferred planet temperature using evolutionary models lies between 160--225~K, which significantly overlaps with the 150--200~K planet temperatures already considered in our analysis. For a planet temperature of 225~K, consistency with the non-detection requires an atmospheric metallicity $= +1.0$~dex in the cloud-free scenario. The optical thickness of water clouds decreases for increasing temperature. Thus, the inclusion of water clouds would not change the requirement of enhanced metallicity (already the case at 200~K, see Figure~\ref{fig:patchy}). Overall, potential systematic uncertainties in the age estimate do not affect the conclusions of this work.

\section{KLIP Parameter Maps}
\label{app:cc}

\begin{figure*}[t]
    \centering
\includegraphics[width=\linewidth]{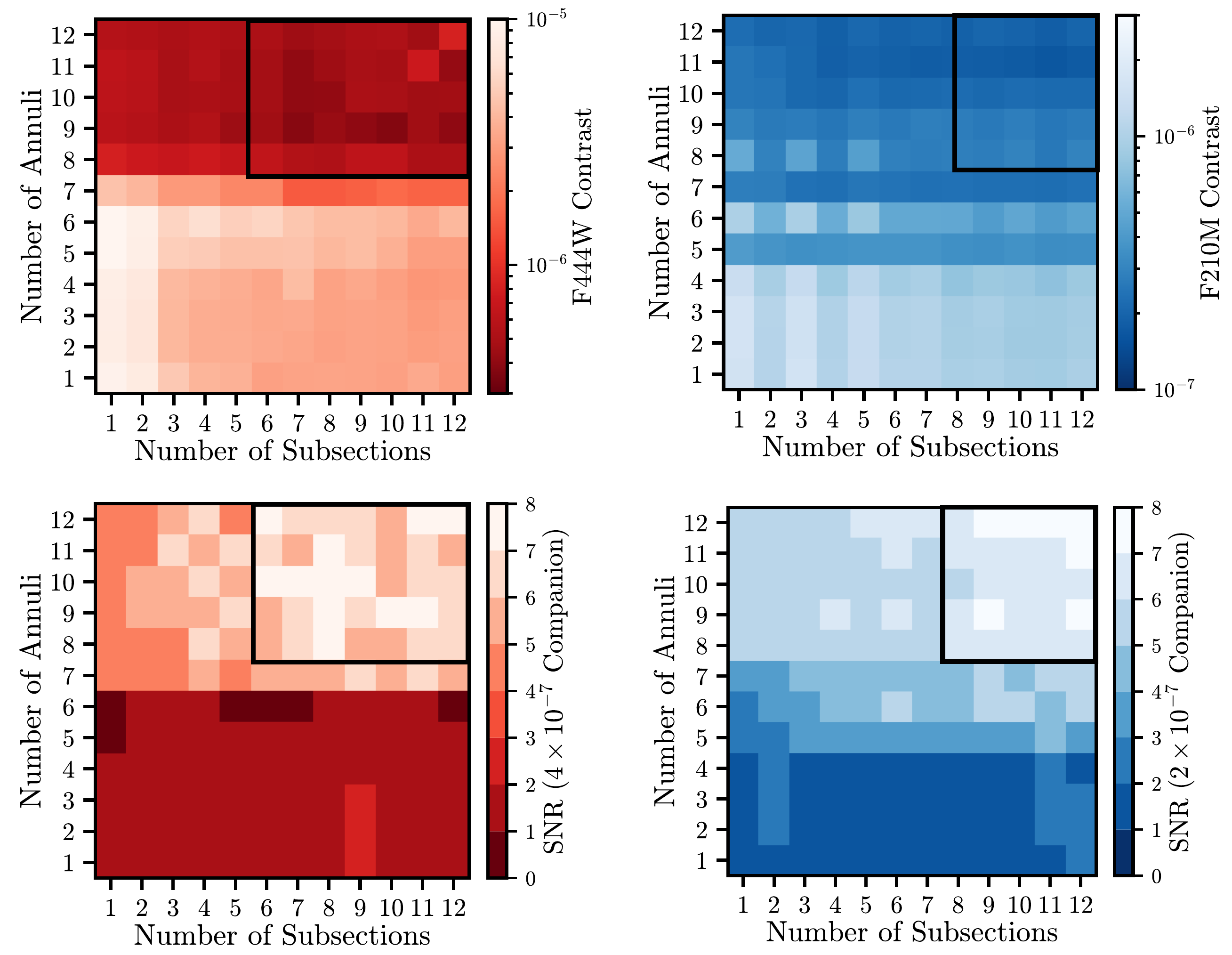}
    \caption{\emph{Top row:} Calibrated F444W and F210M contrast as a function of KLIP reduction parameters. \emph{Bottom row:} Signal-to-noise ratio (SNR) of synthetic companions after injection and recovery at a fixed contrast level, averaged across all position angles of injection, as a function of KLIP reduction parameters. The boxed region in all panels shows the subset of reduction geometries that achieve the best performance. These are considered for the final $5\sigma$ contrast curve calculation.}
    \label{fig:geometry}
\end{figure*}

The goal of this section is to identify the optimal set of KLIP subtraction geometries \citep[e.g.,][]{adams_redai_giant_2023, sanghi_efficiently_2024}, parameterized by \texttt{annulus} and \texttt{subsections} in \texttt{pyKLIP}, for contrast curve calculations (\S\ref{sec:contrast}). We perform the analysis in two steps. 

First, we estimate the contrast performance of our observations for all reduction geometries (and principal components) discussed in \S\ref{sec:data-proc} using the \texttt{spaceKLIP} package following previous work \citep{kammerer_performance_2022, carter_jwst_2023}. We begin by measuring the annular standard deviation of the starlight-subtracted images at a separation of 1\arcsec. The raw contrast, defined as $5\times$ the annular standard deviation, is corrected for small-sample statistics using the Student $t$-distribution following \citet{mawet_fundamental_2014}. The corrected contrast was then calibrated into physical units using a solar metallicity, $T_{\rm eff} = 5100$~K, log~$g = 4.50$~dex (cgs units) BT-NextGen stellar spectrum \citep{allard_model_2011, allard_models_2012} scaled by the interferometrically-determined radius of \epseri \citep[$0.738 \pm 0.003\: R_\odot$,][]{rains_precision_2020}. The coronagraph transmission function was applied to the calibrated contrast. Finally, we calibrate the contrast for KLIP algorithmic throughput via injection-recovery tests conducted at a separation of 1\arcsec\ and at position angles (\texttt{injection\_pas}) $[0^\circ, 330^\circ]$ in steps of 30$^\circ$, excluding position angles overlapping with the expected position of \epserib\ (210$^\circ$, 240$^\circ$). The results are presented in Figure~\ref{fig:geometry} for both filters.

Second, instead of optimizing for the annular contrast at 1\arcsec, we optimize for the signal-to-noise ratio (SNR) of the injected synthetic companions, after PSF subtraction. For each reduction geometry, we inject a $4\times 10^{-7}$ contrast (in F444W) and $2\times 10^{-7}$ contrast (in F210M) companion at a separation of 1\arcsec\ and position angles $[0^\circ, 330^\circ]$ in steps of 30$^\circ$, excluding position angles overlapping with the expected position of \epserib\ (210$^\circ$, 240$^\circ$). The contrast values for injection are selected based on the calibrated annular contrast measured previously. We record the SNR of the synthetic companion \citep{mawet_fundamental_2014} averaged across all position angles of injection. The results are presented in Figure~\ref{fig:geometry} for both filters.

We generally find that using 8--12 annuli divided into 6--12 subsections (and using the maximum number of principal components $= 10$) results in the best annular contrast performance and highest recovered SNR in both filters. It is worth noting that the reduction geometry that yields the best annular contrast does not necessarily yield the highest companion SNR. For our 5$\sigma$ contrast curve calculation in \S\ref{sec:contrast}, we choose to optimize for injected companion SNR since that is the primary metric for signal detection in high-contrast imaging. The boxed parameter combinations in Figure~\ref{fig:geometry} are considered for the final 5$\sigma$ contrast curve calculation.

\bibliography{references.bib,references_EEM.bib}
\bibliographystyle{aasjournalv7}

\end{document}